\documentclass[lettersize,journal]{IEEEtran}
\usepackage{amsmath,amsfonts}
\usepackage{algorithmic}
\usepackage{algorithm}
\usepackage{array}
\usepackage[caption=false,font=normalsize,labelfont=sf,textfont=sf]{subfig}
\usepackage{textcomp}
\usepackage{stfloats}
\usepackage{url}
\usepackage{verbatim}
\usepackage{graphicx}
\usepackage{cite}
\usepackage{balance}
\usepackage{caption}
\usepackage{tabularx}
\usepackage{hyperref}
\usepackage{color,soul}
\usepackage{float}
\usepackage{booktabs}
\usepackage{multirow}
\usepackage{color}
\usepackage{subcaption}
\usepackage[usenames,dvipsnames,svgnames,table]{xcolor}
\usepackage{listings}
\hypersetup{
    colorlinks=true,
    filecolor=magenta,      
    urlcolor=magenta,
}
\usepackage{array}
\usepackage{soul}
\usepackage{xcolor}
\sethlcolor{green}
\newcolumntype{L}{>{\RaggedRight\arraybackslash}m{3cm}}

\begin{document}


\title{Understanding Mental States in Active and Autonomous Driving with EEG}



\author{Prithila Angkan, Paul Hungler, Ali Etemad, \IEEEmembership{Senior Member,~IEEE}
}



\maketitle

\begin{abstract}
Understanding how driver mental states differ between active and autonomous driving is critical for designing safe human–vehicle interfaces. This paper presents the first EEG-based comparison of cognitive load, fatigue, valence, and arousal across the two driving modes. Using data from 31 participants performing identical tasks in both scenarios of three different complexity levels, we analyze temporal patterns, task-complexity effects, and channel-wise activation differences. Our findings show that although both modes evoke similar trends across complexity levels, the intensity of mental states and the underlying neural activation differ substantially, indicating a clear distribution shift between active and autonomous driving. Transfer-learning experiments confirm that models trained on active driving data generalize poorly to autonomous driving and vice versa. 
We attribute this distribution shift primarily to differences in motor engagement and attentional demands between the two driving modes, which lead to distinct spatial and temporal EEG activation patterns. Although autonomous driving results in lower overall cortical activation, participants continue to exhibit measurable fluctuations in cognitive load, fatigue, valence, and arousal associated with readiness to intervene, task-evoked emotional responses, and monotony-related passive fatigue.
These results emphasize the need for scenario-specific data and models when developing next-generation driver monitoring systems for autonomous vehicles.
\end{abstract}


\section{Introduction}
\IEEEPARstart{T}{he} advancement of autonomous driving has impacted the field of human and vehicle interaction. New challenges have developed in understanding driver mental states in autonomous settings, ranging from affect to cognitive load \cite{young2002attention,li2021sensitivity}. As vehicles transition from active to semi- and fully-autonomous modes, drivers experience a shift in their emotional responses and cognitive load \cite{schnebelen2020estimating,merat2019out}. Active driving requires consistent attention, sensorimotor coordination, and real-time decision making throughout the tasks, whereas autonomous driving lacks the motor engagement and continuous decision making.
However, new challenges such as monitoring and readiness to intervene are introduced in autonomous driving scenarios \cite{miller2024navigating}. The transition from active control to passive monitoring alters the neurophysiological engagement during driving, particularly in terms of situational awareness and the ability to resume control when necessary. This impacts attention allocation in the two different driving situations. Understanding these differences in neurophysiological demands is crucial for autonomous driving safety. 


Electroencephalography (EEG) is a powerful tool for investigating mental states such as cognitive load, fatigue, and affect due to its high temporal resolution and ability to capture changes in brain activity \cite{wang2025generative}. 
Non-invasive EEG provides a huge advantage over other forms of monitoring, such as gaze, electrocardiogram (ECG), or electrodermal activity (EDA), due to its direct monitoring of the brain \cite{stuldreher2020physiological,10759719}. 
The portability of modern EEG systems makes them particularly suitable for driving studies, allowing researchers to capture authentic neural responses in driving environments. Recent advances in dry electrode technology have further enhanced the feasibility of EEG-based driving research, enabling high-quality data collection even in the presence of movement artifacts and environmental noise typical of driving scenarios.

Prior EEG studies in the context of driving have focused on fatigue detection \cite{li2024automatic}, vigilance \cite{zheng2017multimodal}, affect \cite{ying2024multimodal}, and cognitive load \cite{angkan2024multimodal} during \textit{\textbf{active}} driving conditions. \textit{However, neural behaviors during autonomous driving due to the lack of motor engagement have not been investigated}. Specifically, a significant gap exists in the literature regarding the neurophysiological differences between active and autonomous driving modes. 
Understanding such differences can eventually lead to the development of safety systems and protocols for autonomous vehicles.


In this paper, we investigate driving under both active and autonomous scenarios using a vehicle simulator. We collect data from 31 participants in each scenario, where 29 participants are common in both the two scenarios. During active driving, the participants maintain full control of the vehicle when completing a series of driving tasks, while during the autonomous scenario, the vehicle operates in self-driving mode and the participant are in the role of passive drivers. In this case, they continuously monitor the road while being prepared to take over the control in unexpected circumstances. We collect EEG data throughout the experiments in order to study the difference in brain activity between the different driving scenarios. The participants also present their mental states in terms of cognitive load, fatigue, valence, and arousal throughout the experiments. We employ a Transformer model to classify the mental states of the drivers in both driving scenarios. Our temporal, complexity-wise, and channel-wise analyses reveal several important findings. First, the temporal pattern analysis reveals higher levels of fatigue during autonomous driving compared to active driving due to the reduced active engagement. Second, we observe that similar mental states are induced across different complexity levels; however, the intensity of the mental states varies between the two scenarios. Third, our topographical analysis identifies scenario-specific brain activation patterns. Finally, the distribution shift between the two scenarios reveals differences in brain activity between active and autonomous driving. 

Our contributions in this paper are summarized as follows:
\begin{itemize}






 \item 
 We present the first EEG-based investigation into the comparison of driver mental states in active versus autonomous driving. Through temporal, complexity-wise, and statistical analyses, we reveal key findings. For instance, participants experience similar patterns of mental states as task complexity increases in both scenarios, but the intensity of these states differs significantly between active and autonomous modes.
 \item We identify a clear distribution shift in EEG data between the two driving modes, which has important implications for transfer learning and model generalization in driver monitoring systems. Our transfer learning experiments confirm this domain shift and demonstrate that mental state monitoring systems trained on active driving data do not generalize well to autonomous contexts. Our findings show the importance of collecting and analyzing data from both scenarios for developing robust monitoring system for driving.
 \item 
 We conduct channel-wise analysis using topographical mapping and statistical comparisons of EEG signals across electrode locations. Our findings reveal that specific brain regions exhibit significantly different activation patterns between active and autonomous driving.
 \item Finally, we address a critical question for autonomous vehicle safety: whether EEG-based mental state classification remains viable during autonomous driving when traditional motor-related neural signatures are absent. Through classification experiments using deep learning, we demonstrate that the absence of motor engagement does not negatively impact the classification, and mental states can be reliably detected from EEG during autonomous driving. This validates the potential for developing dedicated monitoring systems tailored to autonomous driving contexts.
    
\end{itemize}

The rest of this paper is summarized as follows. In Section \ref{section:related_work}, we first provide details on driver's affect and cognitive load studies using EEG. Section \ref{section:experiment_setup} explains the experimental setup, including sensor configurations, driving simulator details, mental state assessments, and data collection protocol. Next, we discuss the data processing 
in section \ref{section:data_processing}. Lastly, in Section \ref{section:results} we provide the details of our analyses, results, and discussions.

\begin{figure*}
    \begin{center}
    \includegraphics[width=1.0\linewidth]{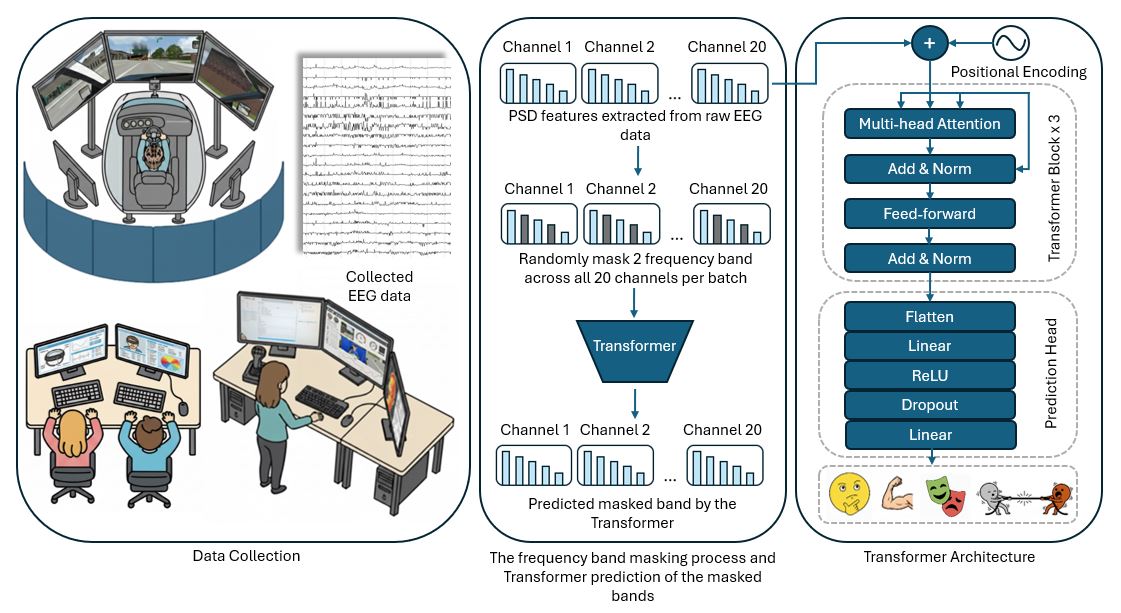}
    \caption{The overall pipeline of our study. }
    \label{fig:pipeline}
    \end{center}
\end{figure*}

\section{Related Work} \label{section:related_work} 

\subsection{Mental State Classification using EEG}

Classification of the mental state of users in different contexts using physiological signals has emerged as an important application spanning from healthcare to human-computer interaction (HCI) \cite{siuly2024exploring,kantipudi2024improved,wu2024grop}. EEG-based mental state classification primarily focuses on identifying cognitive and emotional states such as cognitive load, fatigue, attention levels, and affective dimensions such as valence, arousal, and dominance \cite{pulver2023eeg,chen2024eeg,yousaf2024enhancing}. Cognitive load, defined as the amount of mental effort utilized in working memory during task performance, has been successfully classified using EEG \cite{plass2010cognitive, angkan2024multimodal, pulver2023eeg}. Fatigue detection is another popular research area using EEG signals \cite{li2021temporal,cui2022eeg}. Fatigue is a psychophysiological state of extreme tiredness impacting performance, emotions, and cognition, which can hamper the ability to perform a task \cite{gawron2000overview}. In the context of driving, both peripheral (physical) and central (mental) fatigue can significantly reduce alertness and information retrieval from the memory, increasing the risk of accidents \cite{sikander2018driver}. 
Affective state classification using different physiological signals, including EEG has been widely explored in prior works \cite{liu2024eeg,li2024efficient, zhang2025partial} 
Affective states are typically characterized along multiple dimensions, with the most widely adopted framework being the circumplex model of affect \cite{russell1980circumplex}. This model describes emotions in terms of valence (pleasant-unpleasant), arousal (activated-deactivated) and dominance (not in control-in control). Valence classification focuses on distinguishing between positive and negative emotions. Arousal detection aims to identify the activation level of emotional states, ranging from calm or relaxed to excited or agitated conditions. 


Driving behavior is strongly influenced by a complex set of human factors, including cognitive load, fatigue, attention, and emotional state. When drivers are actively engaged in vehicle control, they must continuously process visual, auditory, and environmental inputs and integrate them into rapid decision-making and motor responses. In active driving, high cognitive load induced by secondary tasks, such as taking on phone, texting, or navigating other devices, can degrade situational awareness and increase the likelihood of errors \cite{engstrom2017effects}. On the other hand, it has been observed that cognitively demanding tasks, due to the nature of the task itself, might increase task performance in expert drivers \cite{engstrom2017effects}. Fatigue during driving further adds to the challenge by reducing vigilance and slowing reaction time, causing impaired executive functions and increasing the risk of accidents \cite{sikander2018driver}. Negative emotional states including stress and frustration can negatively impact driving performance as well. 

\subsection{Driver Mental State Monitoring}

Driver mental state monitoring is a critical safety application in order to ensure road safety and prevent accidents. Drivers experience distinct mental state patterns such as highway hypnosis, fatigue, drowsiness, or high cognitive load \cite{aote2024driver, yang2024video, pulver2023eeg} during monotonous long drives or complex driving situations. Several methods have been used for detecting driver's mental states to ensure road safety ranging from behavioral analysis such as facial video and eye tracking \cite{yang2024video,lu2023jhpfa,li2025glf} to physiological signal analysis such as EEG, ECG, and galvanic skin response (GSR) \cite{fujiwara2023driver,li2024brain,ayoub2023real}.

In \cite{yang2024video}, an effective driver's drowsiness detection method based on facial videos was proposed. They selected the  key facial features and only utilize important information relevant for detecting drowsiness. Another study \cite {lu2023jhpfa}, identified the importance of yawning detection as a sign and driver's fatigue and proposed a network for driver yawning detection. The network combined head pose and facial actions for driver yawning detection using arbitrary poses from videos. Their method consisted of three main components to reduce frame redundancy, generate photo-realistic frontal faces, and fuse head pose and facial features to enhance pose-invariant detection accuracy. In \cite{li2025glf}, a global-local facial spatio-temporal attention fusion method proposed for detecting driver emotions from video data. The framework extracted facial features and applied joint spatial filtering for extracting important information. They processed spatial-temporal data in parallel, used gated fusion for integration, and enhanced emotion class distinction with an angular center loss. In \cite{xiang2025driver}, features from facial expressions and remote photoplethysmography (rPPG) were used from facial video data for multitask driver emotion detection. A cross-modal mutual attention mechanism was used to enhance representations among the different features and calculate their attention. 

ECG signal was used to detect driver's drowsiness in \cite{fujiwara2023driver}. A method was developed to detect unusual changes in the R-R interval which are caused due to drowsiness. Data from 20 participants was collected and a self-attention autoencoder was used for driver drowsiness detection. 
Another study \cite{wang2023ecg} used ECG for driver fatigue detection. Dry electrodes were used to collect ECG data from the driver's palm during driving to subsequently detect fatigue-related characterstics. An ECG based driver's stress detection method was proposed in \cite{wei2024electrocardiogram} to ensure road safety. An Ensembled Multiscale Classifier was proposed that detected drivers stress levels (low, mid, high) by combining fiducial ECG features analyzed with a neural network and non-fiducial ECG features analyzed with a 1-D CNN. 

In \cite{li2024brain}, the relation between EEG and driving decision making was explored. The potential of time-frequency cross-mutual information was used to identify how different brain regions interact during driving tasks. Long short-term memory (LSTM) was used to extract driving related situational embeddings to improve performance. 
In \cite{ayoub2023real}, real-time trust prediction in automated driving settings using physiological signals was explored. GSR, heart rate, and eye-tracking data was employed in this study. A driving simulator was used in this experiment, where the participants were required to comply with the takeover requests (TORs) from autonomous vehicles in three different conditions: control condition, false alarm condition, and miss condition. Finally, changes in their physiological signals were classsified using different machine learning models.

\subsection{EEG Datasets for Driving Contexts}
Several studies have been previously performed for active driving using EEG signals. 
SEED-VIG \cite{zheng2017multimodal} is a publicly available driver's vigilance estimation dataset collected using EEG and Electrooculography (EOG) data. Eye tracking data was also collected in this study, which was used as the ground truth of the physiological measure of vigilance. A sampling rate of 1000 \textit{Hz} was used to collect the EEG data from 12 channels. The data was collected from 23 participants. In this study, the participants drove for 2 hours continuously to induce substantial fatigue. Next, the Sustained-Attention Driving Task dataset \cite{cao2019multichannel} was collected from 27 participants using a 32-channel EEG device with a sampling rate of 500 \textit{Hz}. The data was collected from several different 90-minute sessions, which took place either on the same or different days. The goal of this experiment was to study attention or focus while driving using EEG signals. Here, the car randomly drifted out of the lane, while the participant's job was to notice when it happened. During the experiment, the brain activity was recorded when lane departure occurred and when subjects corrected it. The driver fatigue detection dataset \cite{6f0t-y338-23} was collected from 12 participants while driving a simulated vehicle for 40 to 100 minutes. The data was collected using a 32-channel EEG device with a sampling rate of 1000 \textit{Hz}. The participants had to drive in a simulated highway and a self-reported fatigue questionnaire was used to determine if the participants were fatigued. The PPB-Emo \cite{li2022multimodal} dataset consisted of 3 different experiments with 27, 409, and 40 participants respectively. The first experiment was used to understand the factors that induced driver's emotions and create a questionnaire. In the second experiment, a survey was conducted to identify driving scenarios that evoke specific emotions in drivers, and the results guided the selection of appropriate video-audio stimulus materials. The third experiment utilized the findings from the previous experiments to induce and record EEG emotion data from drivers. Finally, the driver's cognitive load dataset, CL-Drive \cite{angkan2024multimodal} was collected and used to classify the cognitive load of drivers in a simulated vehicle. The data was collected from 21 participants using a 4-channel EEG device with a sampling rate of 256 \textit{Hz}. The participants self-reported their cognitive load scores every 10 minutes during the experiment using the PAAS subjective cognitive load scores consist of 9 levels~\cite{paas1992training}.  
To the best of our knowledge, no prior studies have considered and collected EEG data during autonomous driving scenarios. 


\subsection{EEG-based Driver Mental State Classification}

Recent advances in deep learning have significantly transformed the classification of mental states from EEG in the context of driving. In this section we discuss the prior works where EEG has been used in the context of driving. 

\noindent \textbf{Emotions:}
Driver emotions is an important factor that is known to influence driving performance \cite{gamage2022emotion}. A driver emotion detection method was proposed by \cite{gamage2022emotion}, which classified 4 emotional states of drivers, namely sadness, anger, fearfulness, and calmness. The data was collected from 10 participants and audio and video stimuli were used to induce emotions in the participants at 2-minutes interval. A 5-point Likert scale was used to record the participant's emotions. Another study explored `panic' in the context of driving using graph neural networks (GNN) and attention mechanisms using EEG \cite{chen2024eeg}. The speed of the vehicle was used to induce different levels of panic. In \cite{li2024investigating}, EEG signals and the phase lag index (PLI) were employed to construct brain functional connectivity networks under different emotional states during driving. Graph theory was then applied to analyze these networks, allowing a deeper understanding of the brain's information processing mechanisms. Lastly, \cite{zhang2023electroencephalogram}, proposed a driver emotion detection method using EEG across seven emotions. A manifold learning model was developed to leverage feature extraction on the symmetric positive definite (SPD) matrix manifold.

\noindent \textbf{Fatigue:}
Several studies have been conducted to classify driver distraction in active driving conditions using EEG data \cite{li2021temporal,cui2022eeg}. 
In \cite{li2021temporal}, temporal and spatial information were combined through convolutional layers and gated recurrent units to detect driver fatigue. The data was collected from 24 participants while driving a simulated vehicle during three distracting tasks. 
This work contributed to the understanding of the role of spatial information along with temporal information for EEG signal analysis. 
Another study \cite{cui2022eeg} explored the sample-wise analysis of important EEG features for driver drowsiness recognition. The model leveraged separable convolutions to efficiently capture spatial-temporal patterns of EEG. Another driver fatigue detection study \cite{hu2024eeg} proposed a network to fuse spatial and temporal information along with a strategy to segregate brain regions. Their network consisted of 3 modules, which efficiently captured rich and distinctive long and short-term temporal features. The data was collected from 33 participants while driving for around an hour to induce fatigue. In \cite{gao2023csf}, temporal and spatial frequency domain information were fused for driver fatigue detection. The impact on different frequency bands caused due to fatigue was then explored. The study in \cite{li2024automatic} adopted a neural architecture search approach to find a lightweight and resource limited CNN architecture for real-time fatigue detection.

\noindent \textbf{Vigilance:}
Driver's vigilance is an important factor to consider for road safety. In \cite{zhang2021capsule}, a multimodal method for driver's vigilance estimation was proposed. They used both EEG and EOG signals. To allow the system to concentrate on the most important aspects of the learned multimodal features, they introduced an architecture that integrates a capsule attention mechanism on top of a deep LSTM network. 
In \cite{li2023driver}, the sensitivity of EEG signals to small changes in vigilance was recognized and the potential of using just one frequency band and a small subset of relevant channels was explored for vigilance detection.
In \cite{pan2023residual}, residual attention blocks and a capsule attention mechanism were used for driver vigilance estimation. The method introduced a residual attention network designed to refine the channel-space features of low-level capsules by adaptively rescaling each channel, enabling effective fusion of interdependencies among feature channels. Lastly, \cite{gao2025cross}, bridged the gap between experimental and real-world conditions while maintaining a balance between channel count and detection accuracy. It introduced a framework for EEG-based vigilance detection that supported cross-scenario, cross-subject, and cross-device applications. The work used reduced EEG channel numbers based on commonly shared regions across scenarios.

\noindent \textbf{Cognitive Load:}
Cognitive load plays a critical role in driving performance, directly influencing a driver’s ability to perceive, process, and respond to dynamic road environments \cite{angkan2024multimodal,pulver2023eeg,azizi2024biosignals}. In \cite{pulver2023eeg}, an EEG-based cognitive load classification was proposed. The method utilized a Transformer model and established a foundation for transfer learning between emotion and cognitive load datasets. In \cite{ma2024cognitive}, EEG microstate analysis was applied in a driving simulation based study to assess cognitive load under various phone use and task difficulty conditions. This revealed distinct neural patterns sensitive to the load levels, which could effectively predict safety outcomes. Another study \cite{azizi2024biosignals} leveraged a pre-trained biosignal Transformer (BIOT) for driver cognitive load classification using EEG, along with other physiological signals. The proposed Transformer-based model utilized an encoder-only architecture with efficient self-attention, Fourier-based biosignal tokenization, a perturbation module for augmentation, and hierarchical spatiotemporal embeddings for temporal and spatial encoding. Lastly, \cite{yang2023real}, proposed an attention-enabled recognition network with decision-level fusion to improve driver workload estimation. A multimodal time-series data, including EEG, eye movements, and vehicle states was used in their study.

\section{Experiment Design and Setup} \label{section:experiment_setup}
In this section, we discuss the experimental protocol used in the study. This includes specifics on the setup for the sensors and driving simulator as well as details on the participants, diving scenarios, and mental state assessments.  

\subsection{Overview of Study Design}
In this study, we investigate mental states during active and autonomous driving using a simulator-based controlled study design. 
Participants experienced three complexity levels (low, medium, and high) in each driving scenario, with each level lasting 10 minutes. We collected EEG data using a 20-channel Neuroelectric Enobio device while participants self-reported their cognitive load, fatigue, valence, and arousal every 2 minutes using Likert scales. During the active driving scenario, participants maintained full control of the simulated vehicle, whereas during autonomous sessions, we employed the Wizard of Oz (WOz) setup to simulate autonomous driving behavior. 
Figure \ref{fig:pipeline} shows the entire pipeline from our data collection protocol to mental state classification. The following subsections detail the sensors, driving scenarios, subjective ratings, experimental protocol, and participant details.

\subsection{Sensors}
\begin{figure}[t]
    \centering
    
    \subfloat[The Enobio 20 device.]{%
        \includegraphics[width=0.50\linewidth]{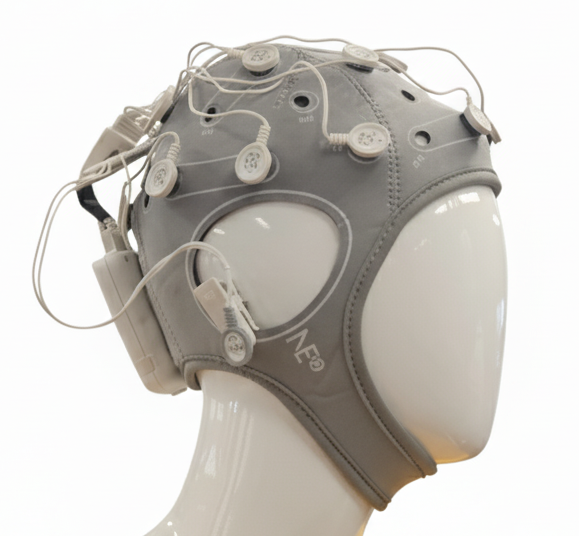}%
    }
    \hfill
    \subfloat[Electrode locations for the Enobio device.]{%
        \includegraphics[width=0.50\linewidth]{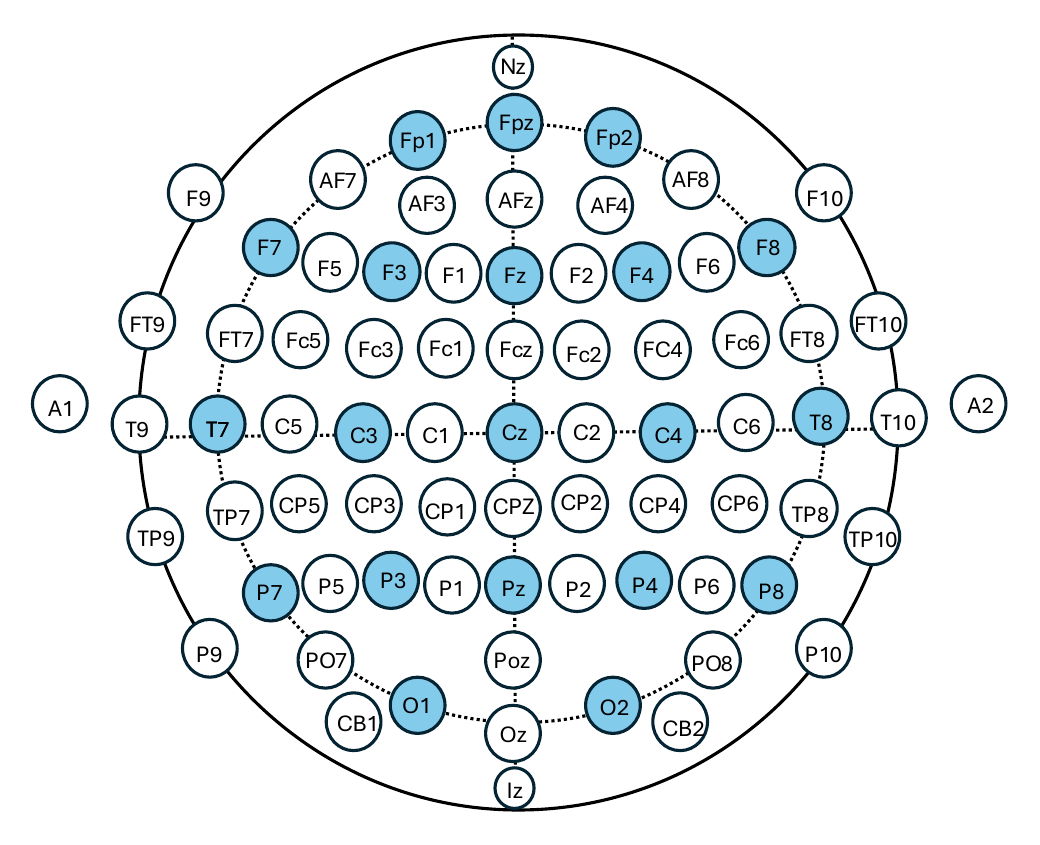}%
    }

    \caption{EEG device used in our study and the electrode location}
    \label{fig:eeg_device}
\end{figure}

The EEG data is collected using the Neuroelectric Enobio 20\footnote{https://www.neuroelectrics.com/products/research/enobio/enobio-20} device shown in Figure \ref{fig:eeg_device} (a). The device has 20 channels: FP1, FP2, F7, F3, Fz, F4, F8, T7, C3, Cz, C4, T8, P7, P3, Pz, P4, P8, O1, O2, and Fpz. These channels are placed according to the international 10-20 system with the help of a cap (Figure \ref{fig:eeg_device} (b)). These electrode positions provide comprehensive coverage of major brain regions, with frontal electrodes (Fp1, Fp2, F3, F4, F7, F8, Fz, Fpz) capturing activity from the prefrontal and frontal cortex areas responsible for attention, cognitive control, action readiness, and interference control \cite{fuster2008prefrontal}. Central electrodes (C3, Cz, C4) are positioned over the sensorimotor cortex to monitor motor-related brain activity \cite{hanselmann2015transcranial}, while temporal electrodes (T7, T8) record activity from regions associated with auditory processing and language functions \cite{bear2020neuroscience}. Parietal electrodes (P3, P4, P7, P8, Pz) capture signals from areas involved in spatial processing, attention, and sensory integration \cite{polich2007updating,corbetta2002control}. Occipital electrodes (O1, O2) monitor visual cortex activity crucial for processing driving-related visual information \cite{eimer2007event}. The device has a sampling rate of 500 \textit{Hz}. To ensure faster preparation time and improved participant comfort during extended driving sessions, we used dry electrodes, which eliminate the need for conductive gel application and reduce setup time. The dry electrode technology also minimizes participant discomfort and allows for more naturalistic experimental conditions.



\subsection{Driving Scenarios}

\begin{figure}
    \begin{center}
    \includegraphics[width=0.9\linewidth]{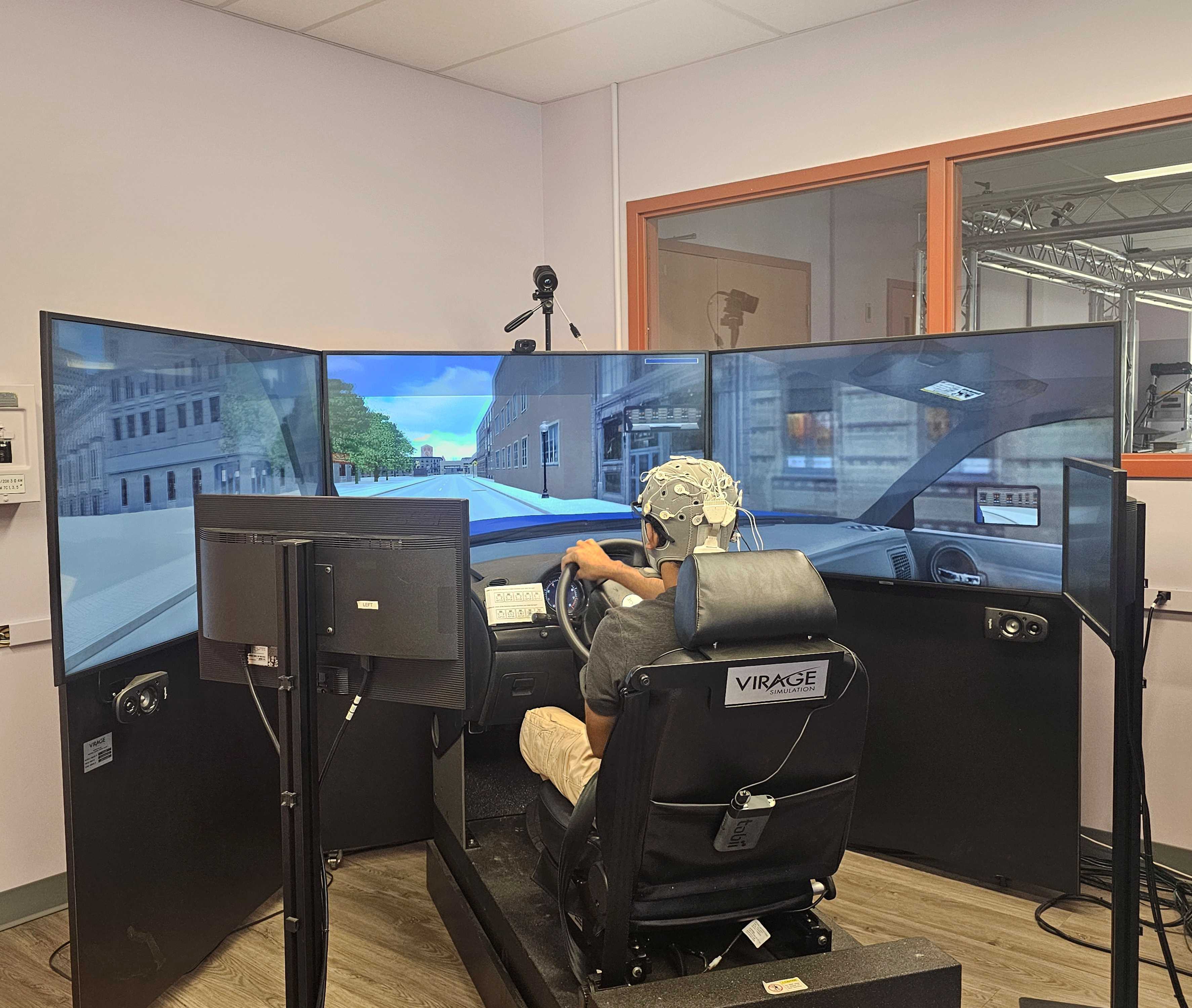}
    \caption{Data collection using the driving simulator.}
    \label{fig:driving_simulator}
    \end{center}
\end{figure}

We use a vehicle simulator to ensure participant safety and enable repeatable and precise experimental control that would be impossible in real-world driving. Simulators eliminate safety risks associated with inducing fatigue and high cognitive load, while allowing exact replication of driving scenarios across all participants. We use the Virage\footnote{https://viragesimulation.com/} driving simulator shown in Figure \ref{fig:driving_simulator}, which provides an immersive environment through realistic visual displays, motions, steering wheel feedback, and environmental audio. The driving scenarios are designed to reflect real-life driving conditions and systematically induce varying levels of cognitive load, fatigue, and emotional responses. Using the settings in the vehicle simulator, we create three distinct driving sessions representing low, medium, and high complexity levels. 
Each complexity level is 10 minutes in length. The low complexity level consists of driving on a mundane road condition on a city highway with a speed limit of 80 \textit{km}/\textit{h}. It is designed for a predictable and easy driving task to establish low cognitive load and emotional responses. The drive is monotonous with minimal traffic and no unexpected events. 
The medium and high complexity scenarios incorporate dynamic urban and highway driving environments at varying speeds (city driving speed of 50 \textit{km}/\textit{h}, highway driving speed of 80 \textit{km}/\textit{h}). This involves realistic traffic conditions with challenging events, which increases cognitive load and emotional responses. Critical driving situations include pedestrian crossings with unpredictable pedestrian behavior, simulated road accidents requiring lane changes and speed adjustments, complex merging maneuvers onto the highways between 2 large trucks that create visibility challenges, emergency braking scenarios triggered by sudden obstacles, and complete stops due to accident-related traffic congestion. To induce negative emotions like frustration, we have incorporated heavy traffic that cannot be overtaken by the participants. 
The distinction between medium and high complexity levels is in the frequency, intensity, and time between these challenging events. Medium complexity scenarios present obstacles with sufficient recovery time between events. However, high complexity scenarios feature more frequent and overlapping stressful events, creating sustained periods of elevated cognitive demand and emotions. Additionally, high complexity scenarios increase the likelihood of poor decision-making due to narrower time windows for safe responses and more aggressive traffic patterns.

\subsection{Participants}
Data was collected from 31 participants in each active and autonomous driving scenario with 29 participants being common among both sessions. Two participants in each scenario did not complete the other scenario due to unavailability. We used posters distributed around campus to recruit participants. Interested individuals were encouraged to provide their availability using the Qualtrics survey tool\footnote{https://www.qualtrics.com/}. 
We maintained all necessary ethical precautions, including a comprehensive informed consent form that detailed the study procedures and time commitment requirements. Prior to data collection, participants were familiarized with all devices and equipment through a practice session to ensure they were comfortable with the experimental setup. Participants were also trained on the subjective rating scales (see Section \ref{section:subjective_ratings}), and a reference sheet was mounted near the dashboard of the simulator for participants to refer to if needed. 
Out of the 31 participants, 17 were male and 14 were female, with ages ranging from 18 to 27 years old and an average of 20.7. 

\subsection{Subjective Ratings}
\label{section:subjective_ratings}
\begin{figure}[htbp]
    \centering

    \subfloat[Cognitive load and fatigue scale.]{
        \includegraphics[width=0.97\linewidth]{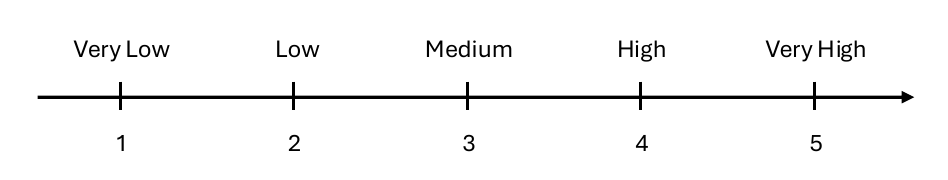}
    }
    
    \subfloat[Valence scale.]{
        \includegraphics[width=0.97\linewidth]{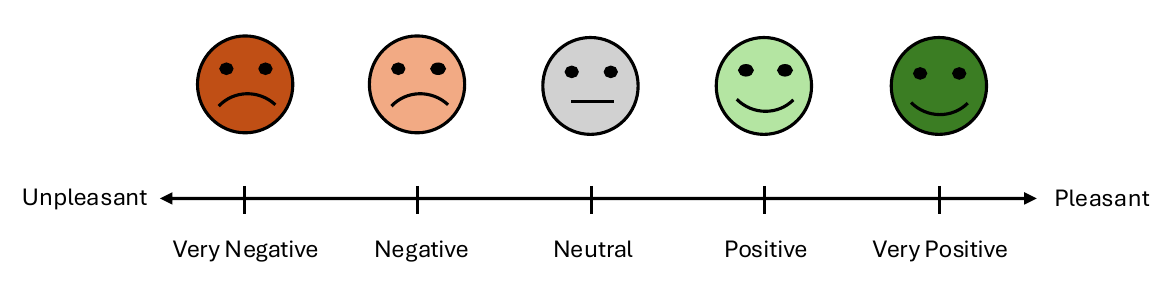}
    }
    
    \subfloat[Arousal scale.]{
        \includegraphics[width=0.97\linewidth]{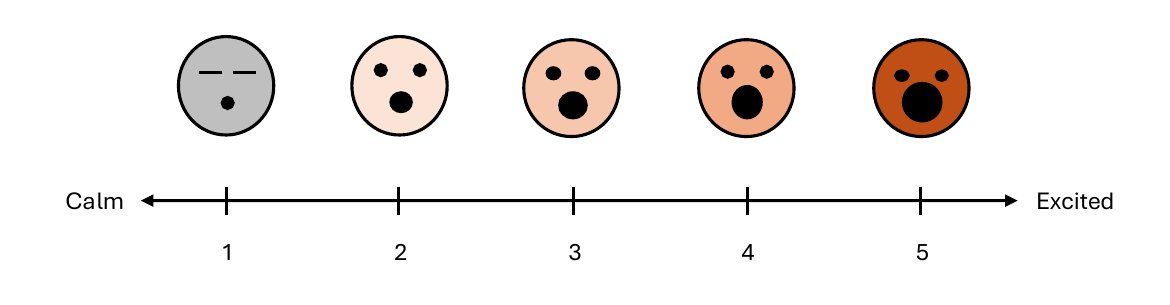}
    }
    
    \caption{Self-reporting scale of cognitive load, fatigue and affect showing (a) Cognitive load and fatigue ranging from very low to very high (b) valence scale representing emotional pleasantness from very negative to very positive, and (c) arousal scale representing emotional activation from very low (calm) to very high (excited).}
    \label{fig:self_reporting_scale}
\end{figure}

Participant's self-reported cognitive load, fatigue, valence and arousal scores as the ground-truth in this study. We have used a 5-point Likert scale for the mental states shown in Figure \ref{fig:self_reporting_scale}. Every 2 minutes during the experiments, participants are asked to report their mental states using the scale ranging from ``Very Low'' to ``Very High'' for cognitive load and fatigue (Figure \ref{fig:self_reporting_scale}(a)), ``Very Negative'' to ``Very Positive'' for valence (Figure \ref{fig:self_reporting_scale}(b)) and ``1'' being very calm to ``5'' being very excited for arousal (Figure \ref{fig:self_reporting_scale}(c)). To monitor for possible simulation adaptation syndrome (SAS) throughout the study, participants also reported their level of nausea on a scale of 0 to 5, 0 being ``None'' and 5 being ``Very High''. At any point during the experiment, if a participant experienced nausea of level of 2 or higher, we stopped the experiment immediately and their data was excluded from the study. 

\subsection{Experiment Protocol}
\begin{figure}
    \begin{center}
    \includegraphics[width=0.9\linewidth]{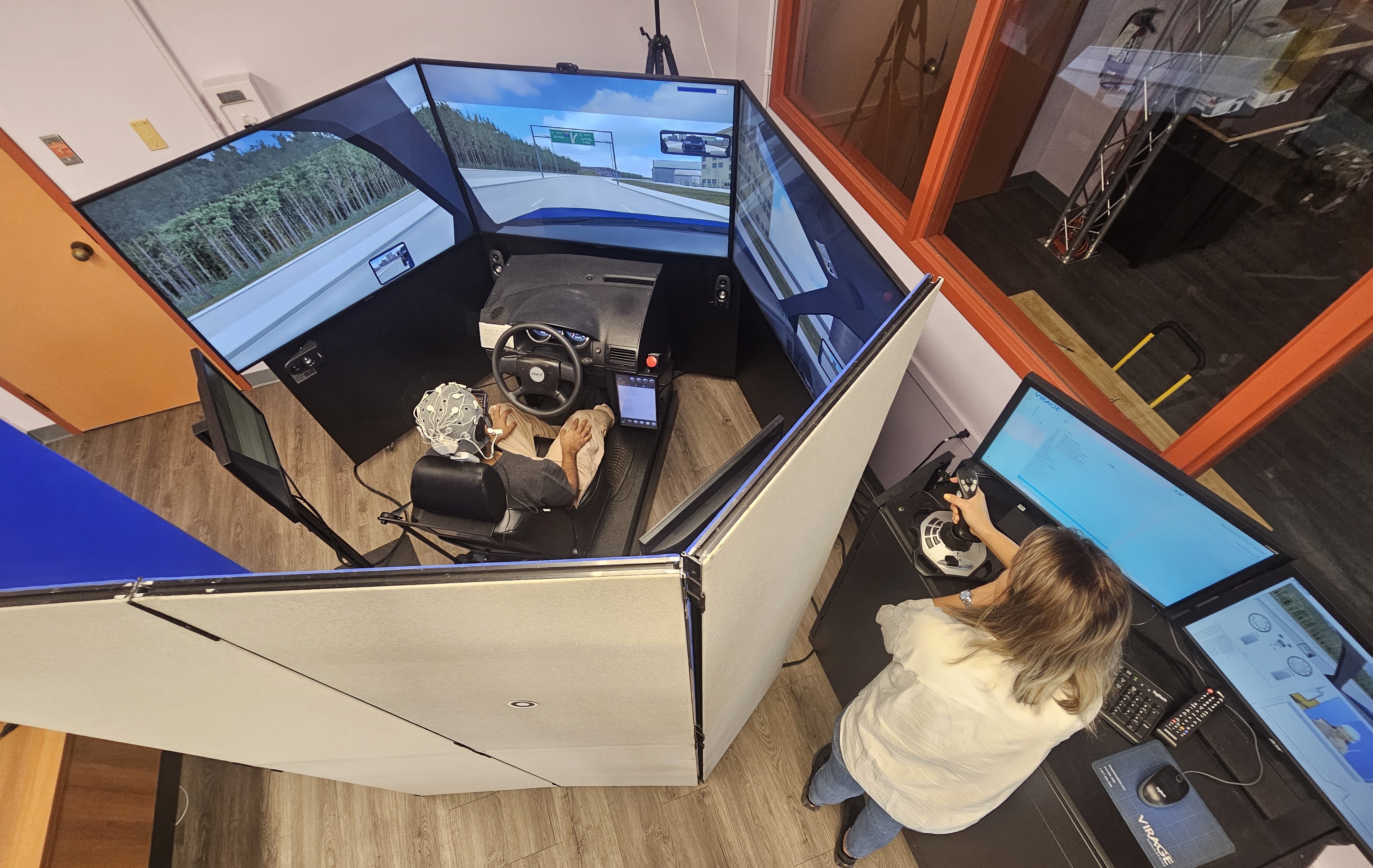}
    \caption{The Wizard of Oz setup used in our study: a hidden human operator controls the vehicle while participants experience simulated autonomous driving.}
    \label{fig:woz}
    \end{center}
\end{figure}

Participants were informed about the experimental procedures before the simulation, and written consent was obtained. Ethics approval for the study was granted by the General Research Ethics Board (GREB) at Queen’s University. Each participant was required to complete two driving scenarios (active and autonomous) on two different days. Each scenario lasted 60 to 90 minutes in total. Participants were asked to fill out a pre-study questionnaire and consent form before the experiment. 
To reduce SAS, a baseline driving task was completed by each participant. 
The participants were given 5 minutes of rest between each complexity scenario. The scenarios high, medium, and low were selected randomly. During the active scenario, the participants drove the simulated vehicle themselves. During the autonomous scenario, we used the WOz setup, where a human operator drives the simulated vehicle using a joystick while hiding from the participant behind a partition, as shown in Figure \ref{fig:woz}. The WOz setup has been previously used in several studies where fully autonomous technology is underdeveloped \cite{sperrle2024wizard,gu2024data}. The hidden operator went through several hours of practice driving sessions to become comfortable driving the vehicle simulator with a joystick. The goal of this study is to understand the difference between the two driving scenarios and investigate whether EEG data collected during autonomous driving contains enough information for mental state classification even without active motor engagement. 

\section{Data processing} \label{section:data_processing}

In this section, we explain the processing pipeline for the data acquired in our study. The goal is two-fold. Our first goal is to evaluate how machine learning models perform on autonomous driving scenarios where active engagement is absent. Second, we intend to use the learned representations and classification performance of the two scenarios, active and autonomous, to determine whether a qualitative and quantitative distribution shift exists among the two scenario types. Following, we describe the data processing procedures.

\subsection{Pre-processing}
We adopt a pre-processing pipeline similar to that in \cite{pulver2023eeg,angkan2024multimodal}. To mitigate unwanted artifacts and environmental noise, the raw EEG signals are first passed through a second-order Butterworth bandpass filter with a passband frequency range of 1–75 \textit{Hz}. In addition, a notch filter with a quality factor of 30 is applied at 60 Hz to suppress power line interference. After filtering, the continuous signals are segmented into 10-second segments to capture short-term temporal dynamics \cite{angkan2024multimodal}. Each segment is then decomposed into five known EEG frequency bands Delta (1–4 \textit{Hz}), Theta (4–8 \textit{Hz}), Alpha (8–12 \textit{Hz}), Beta (12–31 \textit{Hz}), and Gamma (31–75 \textit{Hz}). 

We extract absolute power spectral density (PSD) features for each frequency band, following prior studies \cite{pulver2023eeg, angkan2024multimodal, bhatticlare2025}. PSD quantifies how the signal’s power is distributed across frequencies, offering a robust representation of EEG neural activity. To compute PSD, we employ Welch’s method \cite{solomon1991psd}, where the EEG signal is divided into smaller overlapping sections. Discrete Fourier Transform (DFT) is computed for each section, and the squared magnitudes are averaged. This approach reduces variance and produces smoother, more reliable spectral features. Finally, the extracted PSD values are standardized using z-score normalization \cite{dai2021electroencephalogram}, ensuring that all features have a comparable scale and are suitable for subsequent deep learning analysis.

\subsection{Classifier}
\label{section:classifier}
Transformer networks have revolutionized machine learning across multiple domains, from natural language processing and computer vision to physiological signal analysis \cite{wang2024eegpt,feng2025dit4edit,wu2024llm}. They can capture long-range dependencies and complex relationships through self-attention mechanisms \cite{vaswani2017attention}. Originally introduced for sequence-to-sequence tasks in language modeling, Transformers have proven exceptionally effective at learning contextual relationships between elements in sequential data like EEG. The multi-head self-attention mechanism enables each element to look at other elements simultaneously to capture important patterns. For EEG, the relation between different features, bands and spatial locations are valuable. Therefore, for our study, we have implemented a Transformer network. 

Our Transformer network processes data from 20 channels across 5 frequency bands. The data is flattened from the original (batch\_size, 20, 5) format into a 100-element sequence (batch\_size, 100), where each element represents a unique channel-frequency combination. The architecture starts with an input projection layer of shape (1,64) that embeds each feature into a 64-dimensional space (batch\_size, 100, 64), followed by learnable positional encodings of size (1, 100, 64) to preserve spatial information.
There are 3 Transformer encoder layers with a 4-head self-attention mechanism with query, key, and value projection matrices of dimensions (64, 64). The feed-forward networks project to 256 and 64 dimensions (batch\_size, 100, 64). The classification head comprises a linear layer of size 128 followed by batch normalization, ReLU activation, dropout (0.25), and a final linear layer of size 1 followed by Sigmoid activation for binary output. The 4-head attention in each of the 3 Transformer layers creates a 100 $\times$ 100 attention matrix that captures relationships between different brain regions and frequency bands. After processing through the Transformer layers, global average pooling combines information from all 100 features (batch\_size, 100, 64) into a single 64-dimensional vector (batch\_size, 64), which is then fed to the classification head. The Transformer architecture is shown in Figure \ref{fig:pipeline}. We train our Transformer classifier for 50 epochs using the Adam optimizer with a learning rate of 0.0001 and binary cross-entropy with logits as the loss function for binary classification. A batch size of 64 is used for training. 


\subsection{Training scheme}
We employ a Leave-One-Subject-Out (LOSO) cross-validation approach for training, where the model is trained on data from all participants except one, which serves as the test subject. We repeat this process to test on all the participants and report the average of the accuracies and F1 scores. 
We use binary evaluation for our study. We divide each subjective rating into two groups ``high'' and ``low'', where ``high'' contains values 3, 4, and 5, while ``low'' contains values 1 and 2 for all the mental states.

\subsection{Self-supervised pretraining} \label{section:pretraining}
To allow the Transformer model to learn generalized EEG features, we implement a pretraining strategy using a self-supervised masked frequency band reconstruction task. Pretraining allows models to learn generalizable representations from large unlabeled data before fine-tuning on specific tasks. This often leads to improved performance. In the context of EEG analysis, pretraining on EEG datasets would enable the model to capture universal patterns of brain activity that can transfer across different tasks. The pretraining process is shown in Figure \ref{fig:pipeline}

\subsubsection{Pretraining strategy}
For pretraining, we use 2 EEG datasets. The pretraining approach employs a Transformer model with an identical architecture to our main classifier, but with a modified output head that reconstructs the original 100-dimensional input (20 channels $\times$ 5 frequency bands) rather than producing binary outputs. During pretraining, the model learns to reconstruct complete EEG spectral patterns from masked inputs where 2 out of 5 frequency bands are randomly set to zero. This forces the Transformer to learn meaningful representations of cross-channel and cross-frequency relationships. This masking strategy operates at the frequency band level, where each band spans every 5th index in the flattened feature vector. The model learns to predict missing spectral information based on spatial and remaining frequency contexts. We select the same 20 electrode positions from pretraining datasets that are used in our main classification task and apply identical pre-processing steps, including PSD computation and normalization to maintain consistency between pretraining and fine-tuning phases. We train the model for 200 epochs using L1 loss to minimize reconstruction error between predicted and original frequency band values. Adam optimizer and a learning rate of 0.0001 are used.

\subsubsection{Pretraining dataset}
We use the popular SEED \cite{zheng2015investigating,duan2013differential} and SEED-IV \cite{zheng2018emotionmeter} datasets for our pretraining. The \textbf{SEED} \cite{zheng2015investigating,duan2013differential} dataset contains EEG recordings collected while the participants watched 15 film clips of 4 minutes in length. Each clip is labeled as positive, negative, or neutral. The data was collected from 15 participants, among whom there were 7 males and 8 females with an average age of 23.27. The data was collected in 2 sessions and the EEG signals were recorded using a 62-channel EEG device at a sampling rate of 1000 \textit{Hz}, which was then downsampled to 200 \textit{Hz}. 
\textbf{SEED-IV} \cite{zheng2018emotionmeter} dataset includes EEG recordings from 15 participants (7 males and 8 females) with an age range of 20-24. The participants watched 72 film clips categorized into 4 emotions: happy, sad, fear, and neutral. Three sessions were conducted on separate days where each participant watched 24 clips per session. EEG signals were captured with 62 channels at 1000 \textit{Hz} and subsequently downsampled to 200 \textit{Hz}. 

\subsubsection{Downstream strategy}
For the downstream task of classifying the mental states, we use two distinct approaches: partial fine-tuning and full fine-tuning. In the \textbf{partial fine-tuning} approach, the pretrained Transformer encoder layers are loaded with the pretrained weights. Their parameters are frozen, meaning they cannot be updated during training. Only the classification head (the MLP layers that output binary predictions) is randomly initialized and trained from scratch with our own data. 
In the \textbf{full fine-tuning} approach, the entire model architecture is initialized with the pretrained weights from the reconstruction task, but all parameters remain trainable during the downstream classification training. This approach typically yields better performance than partial fine-tuning feature extraction since the model can specialize its representations for the target task.
For both partial and full fine-tuning, we use an optimizer, learning rate, loss, epoch and batch size described earlier in \ref{section:classifier}.



\section{Results and Discussions} \label{section:results}
In this section, we present the results of our experiments and our findings. 

\subsection{Mental states over time}
\begin{figure}[t]
    \centering
    
    \subfloat[]{%
        \includegraphics[width=0.50\linewidth]{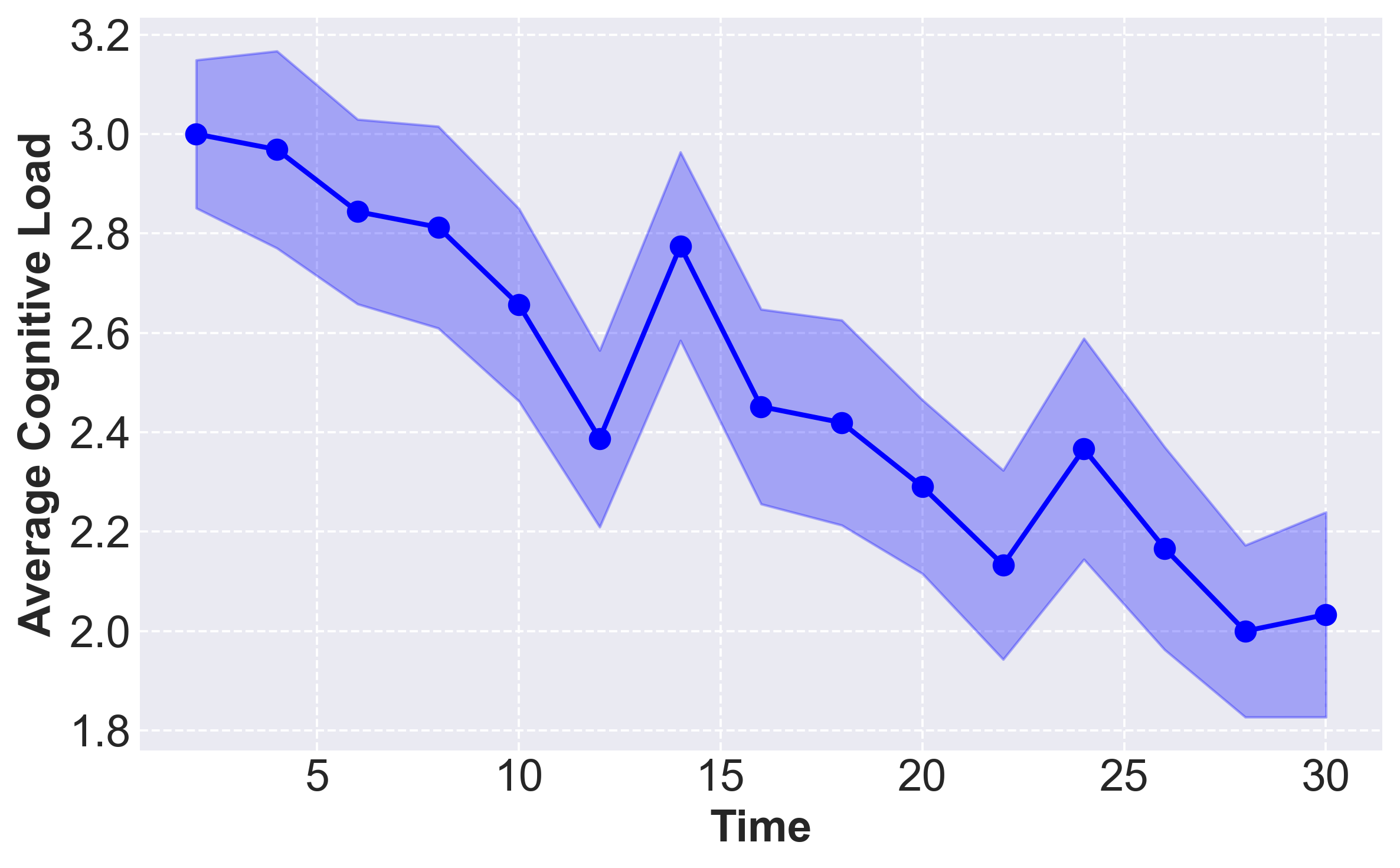}%
    }
    \hfill
    \subfloat[]{%
        \includegraphics[width=0.50\linewidth]{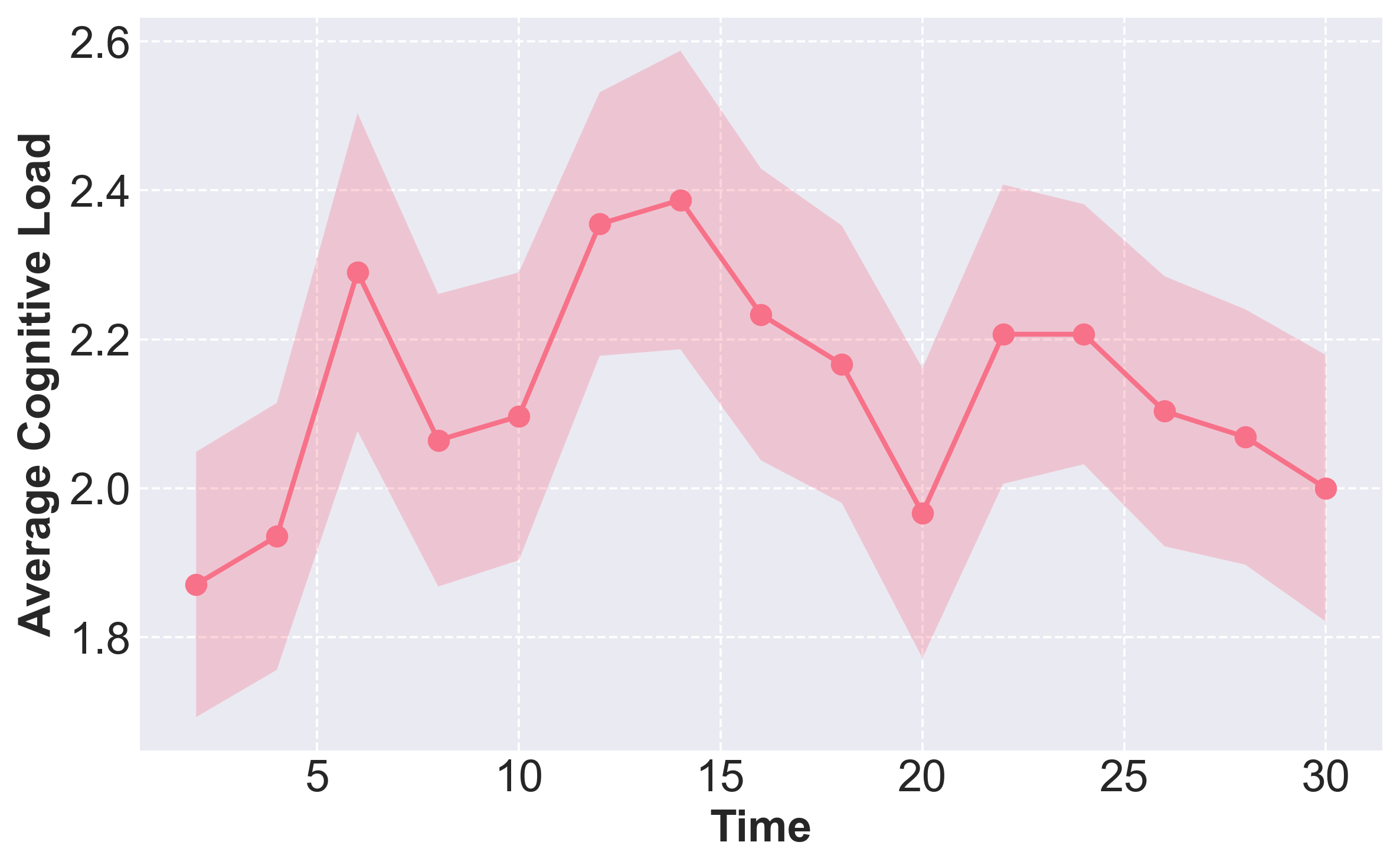}%
    }
    
    \subfloat[]{%
        \includegraphics[width=0.50\linewidth]{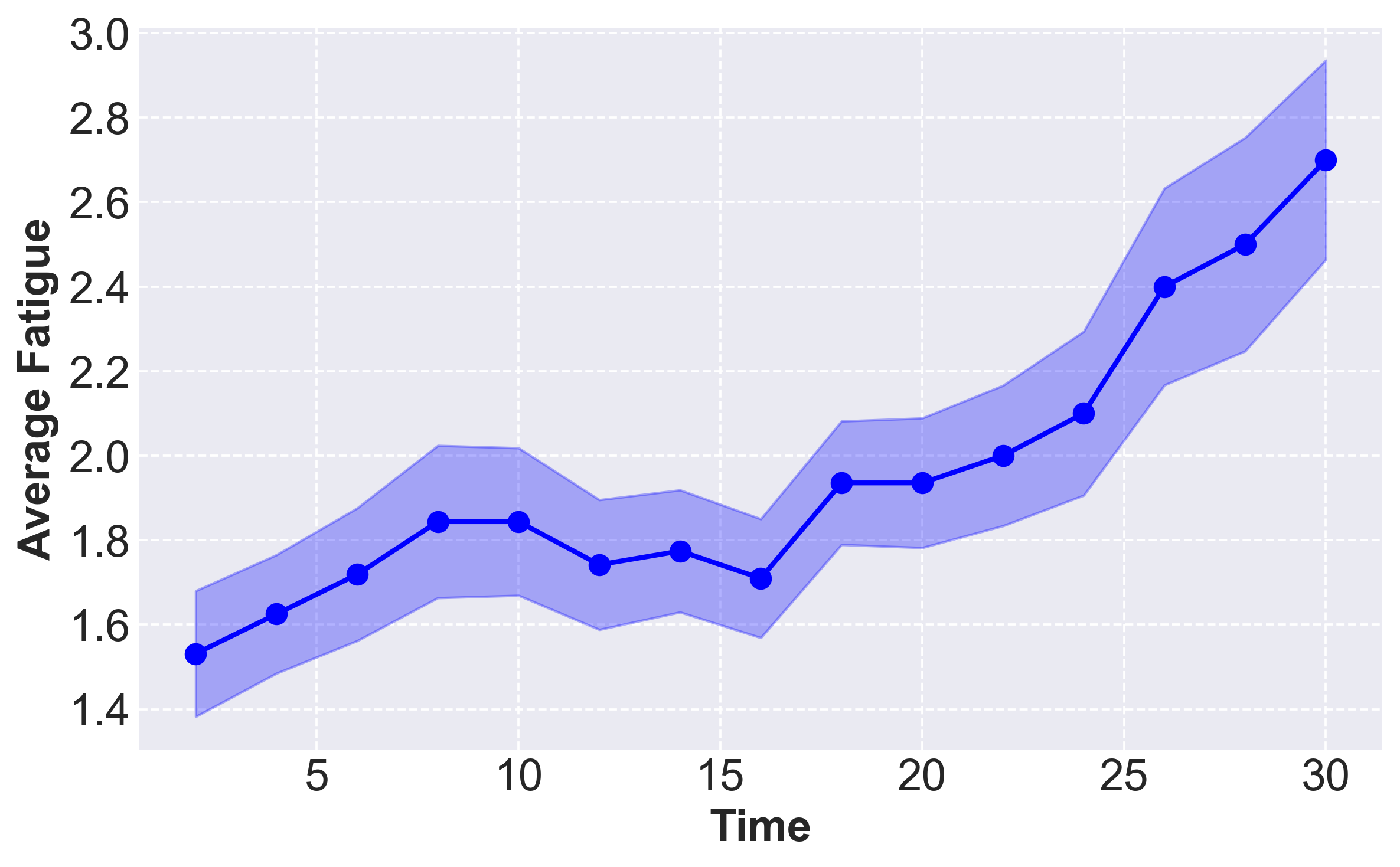}%
    }
    \hfill
    \subfloat[]{%
        \includegraphics[width=0.50\linewidth]{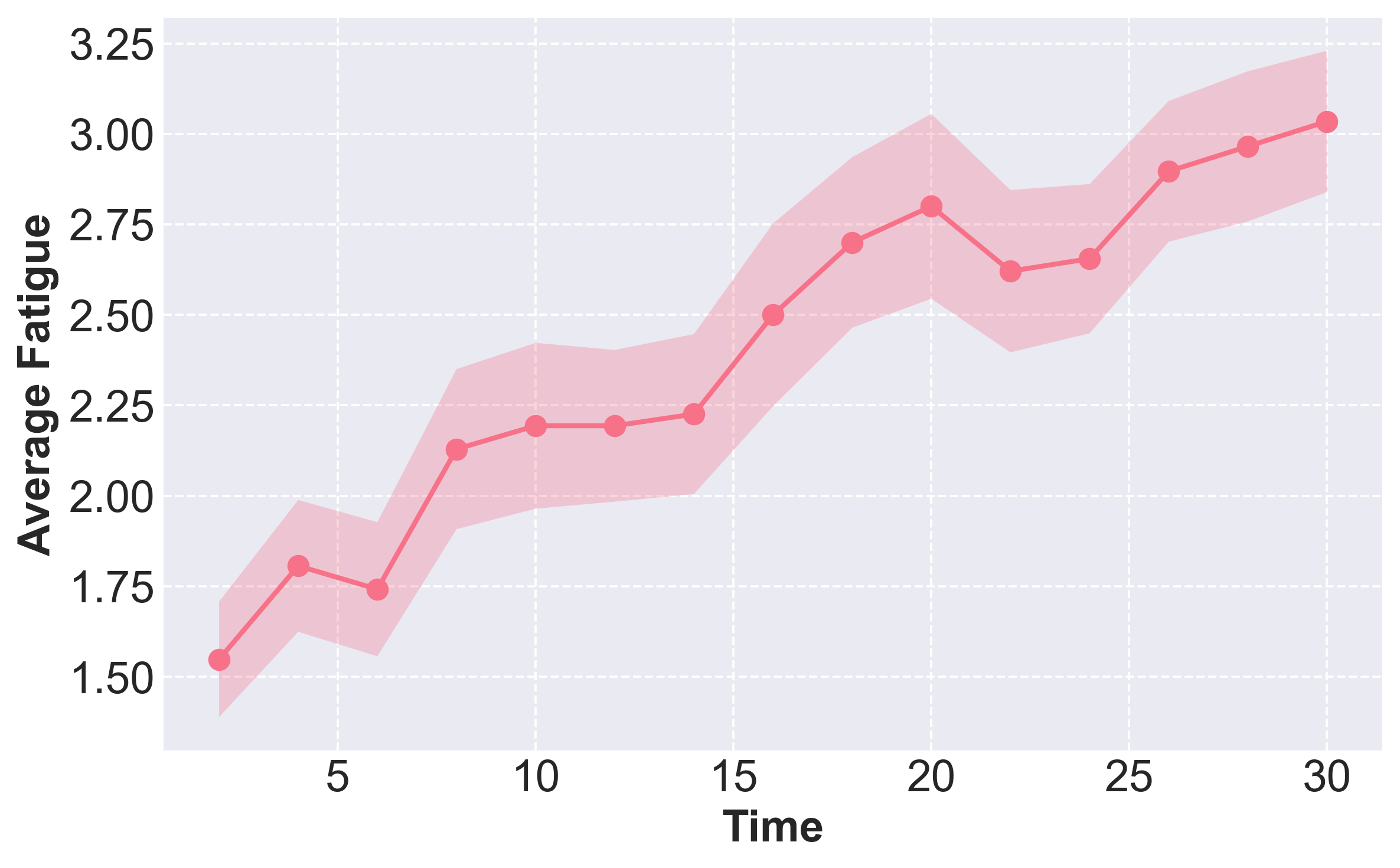}%
    }
    
    \subfloat[]{%
        \includegraphics[width=0.50\linewidth]{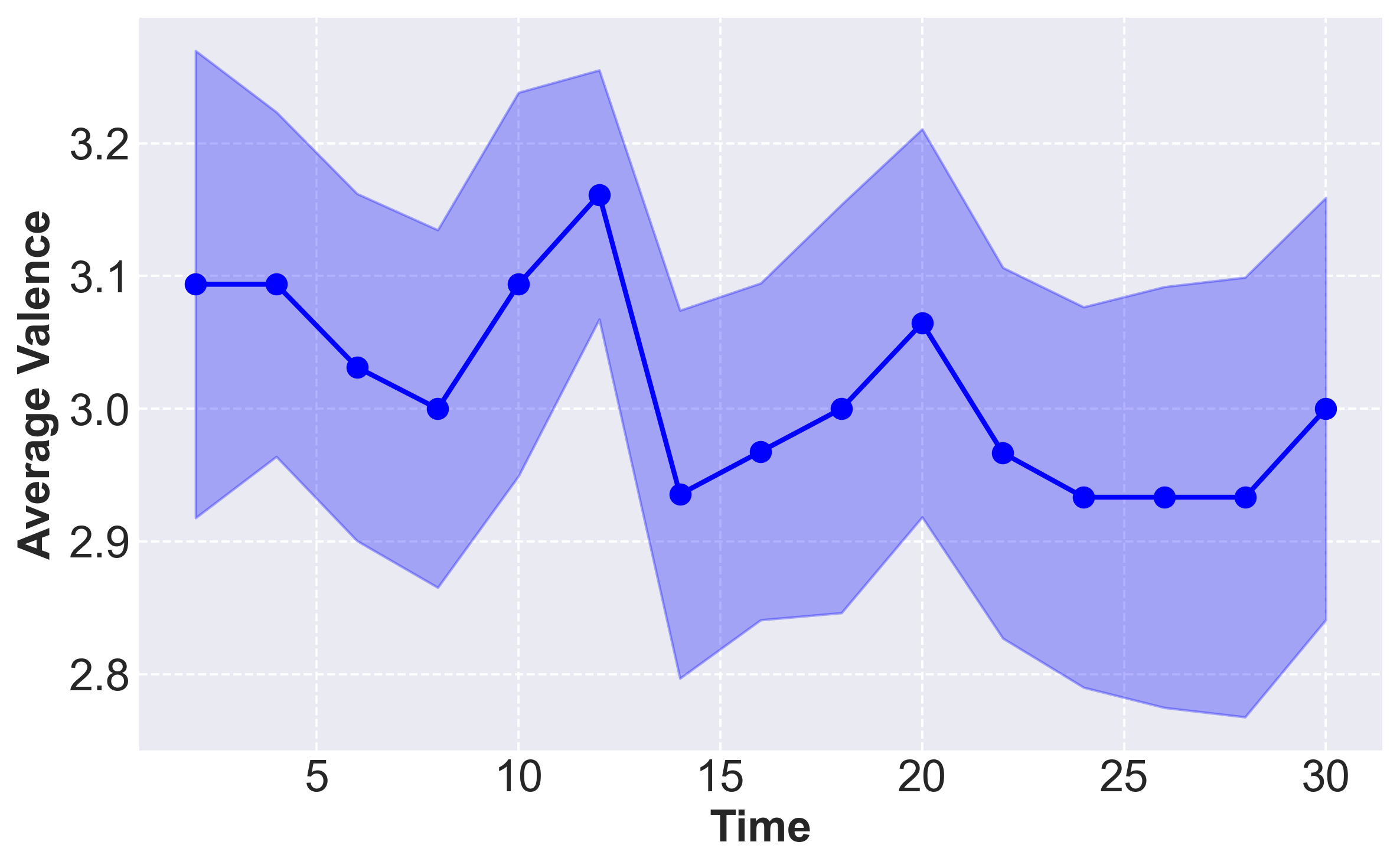}%
    }
    \hfill
    \subfloat[]{%
        \includegraphics[width=0.50\linewidth]{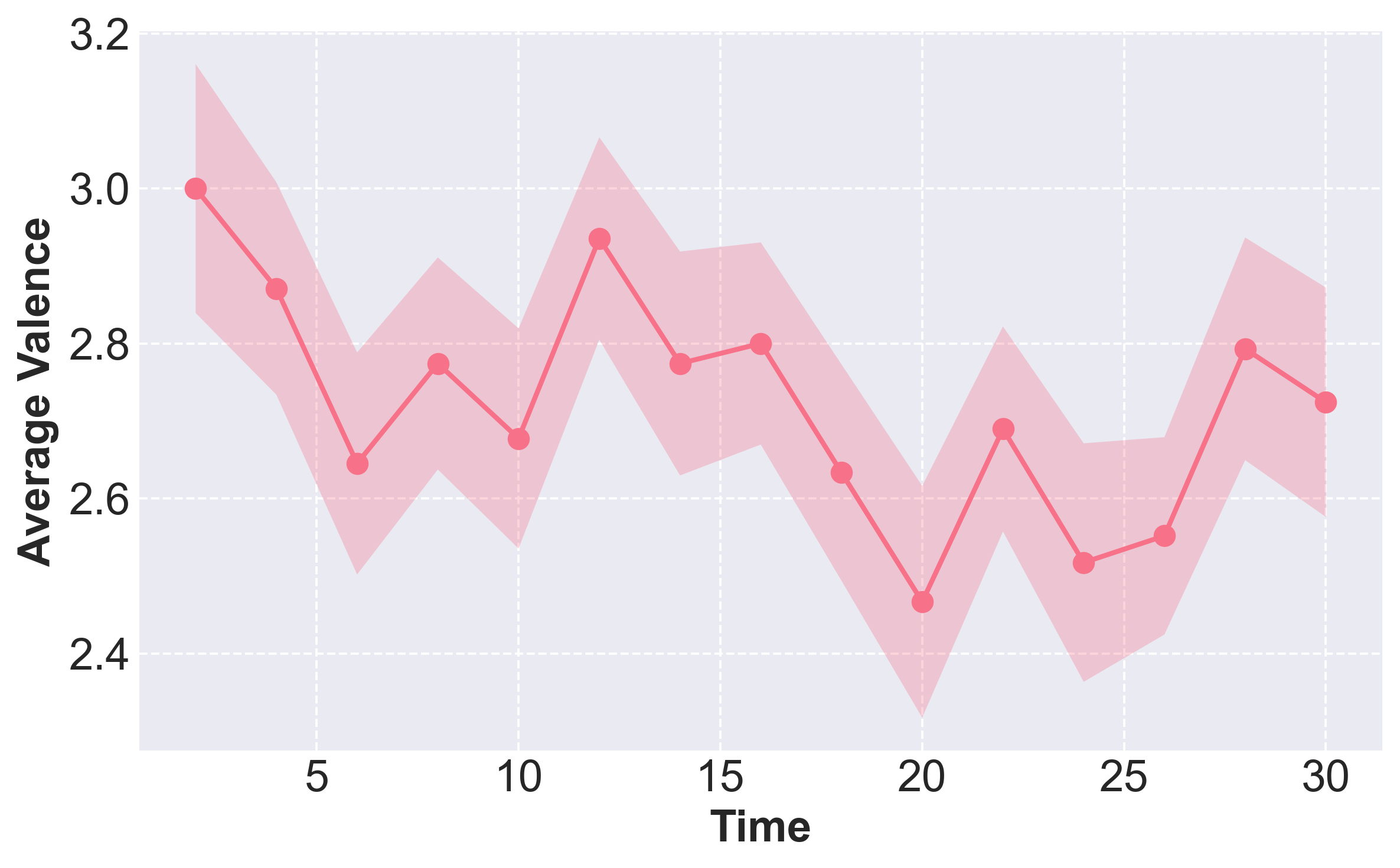}%
    }
    
    \subfloat[]{%
        \includegraphics[width=0.50\linewidth]{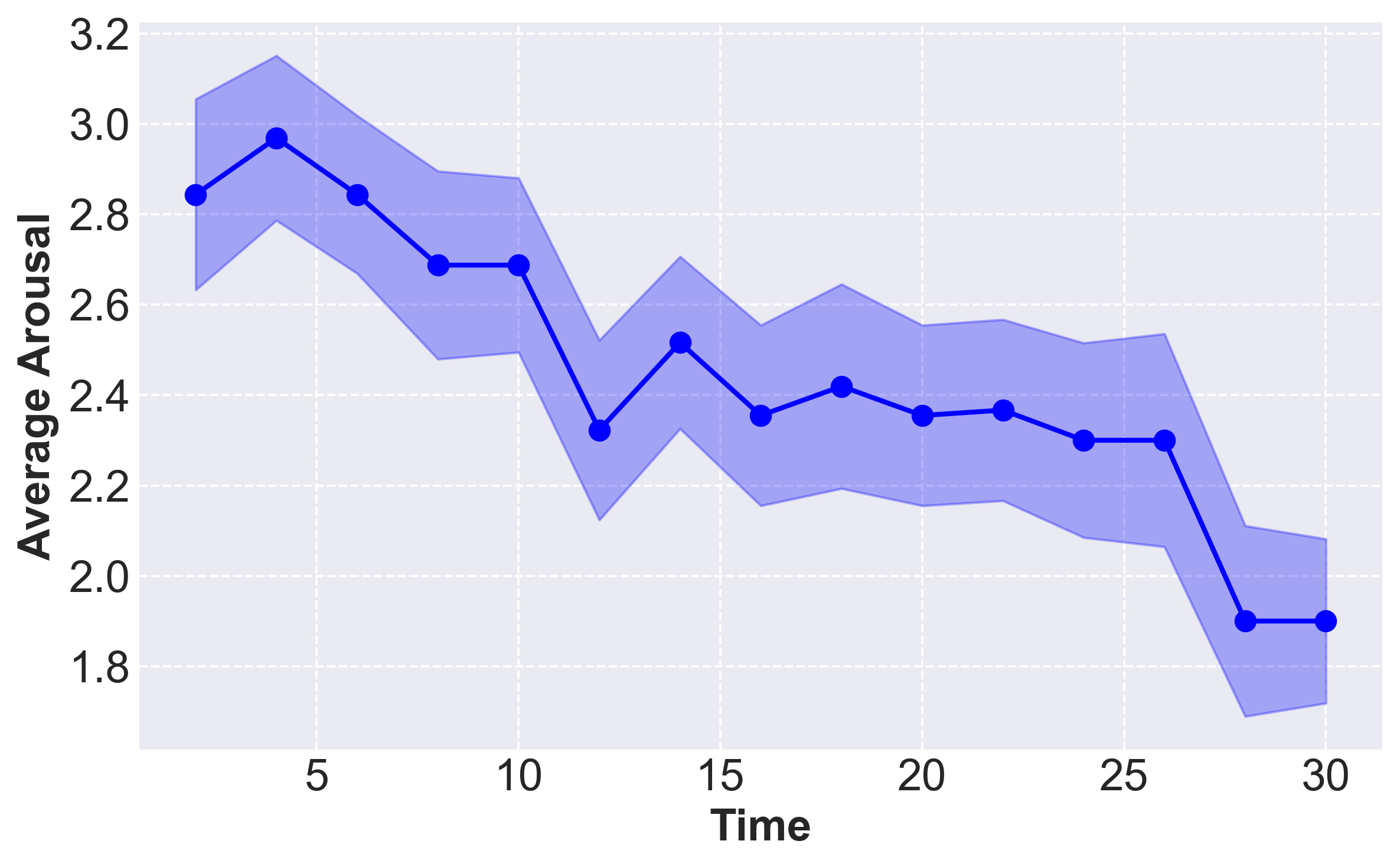}%
    }
    \hfill
    \subfloat[]{%
        \includegraphics[width=0.50\linewidth]{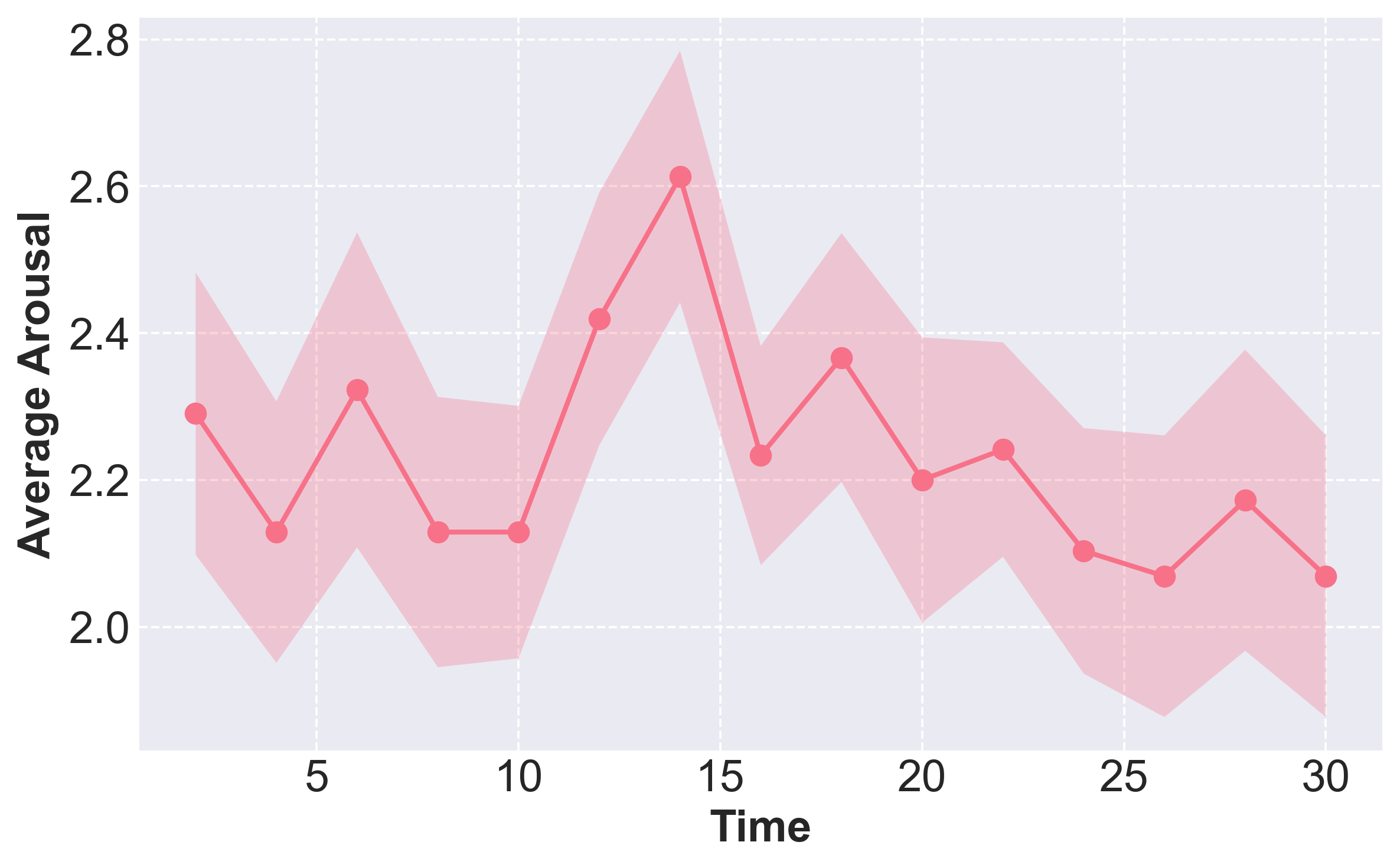}%
    }
    
    \caption{Mental states over time. The left plots are for active scenarios while the right plots are for autonomous scenarios.}
    \label{fig:mental_states_over_time}
\end{figure}

In this section, we demonstrate the relation between each mental state over time. The plots in Figure \ref{fig:mental_states_over_time} illustrate the mental states over time of the experiment, irrespective of the complexity levels. We also show the standard error (SE) in the graphs, which is calculated by the standard deviation (std) divided by the square root of the sample size (n), $SE = \frac{std}{\sqrt{n}}$.

\noindent \textbf{Cognitive Load:} Figure \ref{fig:mental_states_over_time} (a) active and (b) autonomous, show the average cognitive load experienced by the participants over time. In active scenario, we can see that the cognitive load gradually decreases as the participants get used to the driving task. For autonomous scenario, the cognitive load fluctuates due to the randomness of events and attention requirement. 

\noindent \textbf{Fatigue:} The average fatigue progression plots across all participants in Figure \ref{fig:mental_states_over_time} (c) active and (d) autonomous scenarios, reveal that over time, the experienced level of fatigue increases. During active session (c), the participants show a relatively gradual increase in fatigue levels starting at around 1.3 and reaching to approximately 2.7 by the end of 30 minutes. In contrast, autonomous driving (d), demonstrates a steeper fatigue pattern at the beginning, also starting at around 1.3 and ending around 3.0. This pattern suggests that, active driving produces more sustained cognitive engagement while autonomous driving may involve periods of heightened vigilance and boredom-induced fatigue that can fluctuate as participants alternate between passive monitoring and mind-wandering in an autonomous scenario.

\noindent \textbf{Valence and Arousal:} The valence and arousal over time are shown in Figure \ref{fig:mental_states_over_time}, where (e) and (f) are valence, and (g) and (h) are arousal for active and autonomous scenarios, respectively. For both valence and arousal, we can see random peaks throughout the entire time, which is caused by events that induced certain emotions in the participants. 
The downward curve for active driving arousal suggests that as participants become more accustomed to operating the simulated vehicle, their arousal naturally decreases. In contrast, this pattern does not appear in the autonomous driving scenario. Because participants must establish trust in the autonomous system, which is inherently challenging, their arousal fluctuates whenever events occur rather than steadily declining.




    

\subsection{Complexities vs. mental states}
\begin{figure}[t]
    \centering
    
    \subfloat[]{%
        \includegraphics[width=0.50\linewidth]{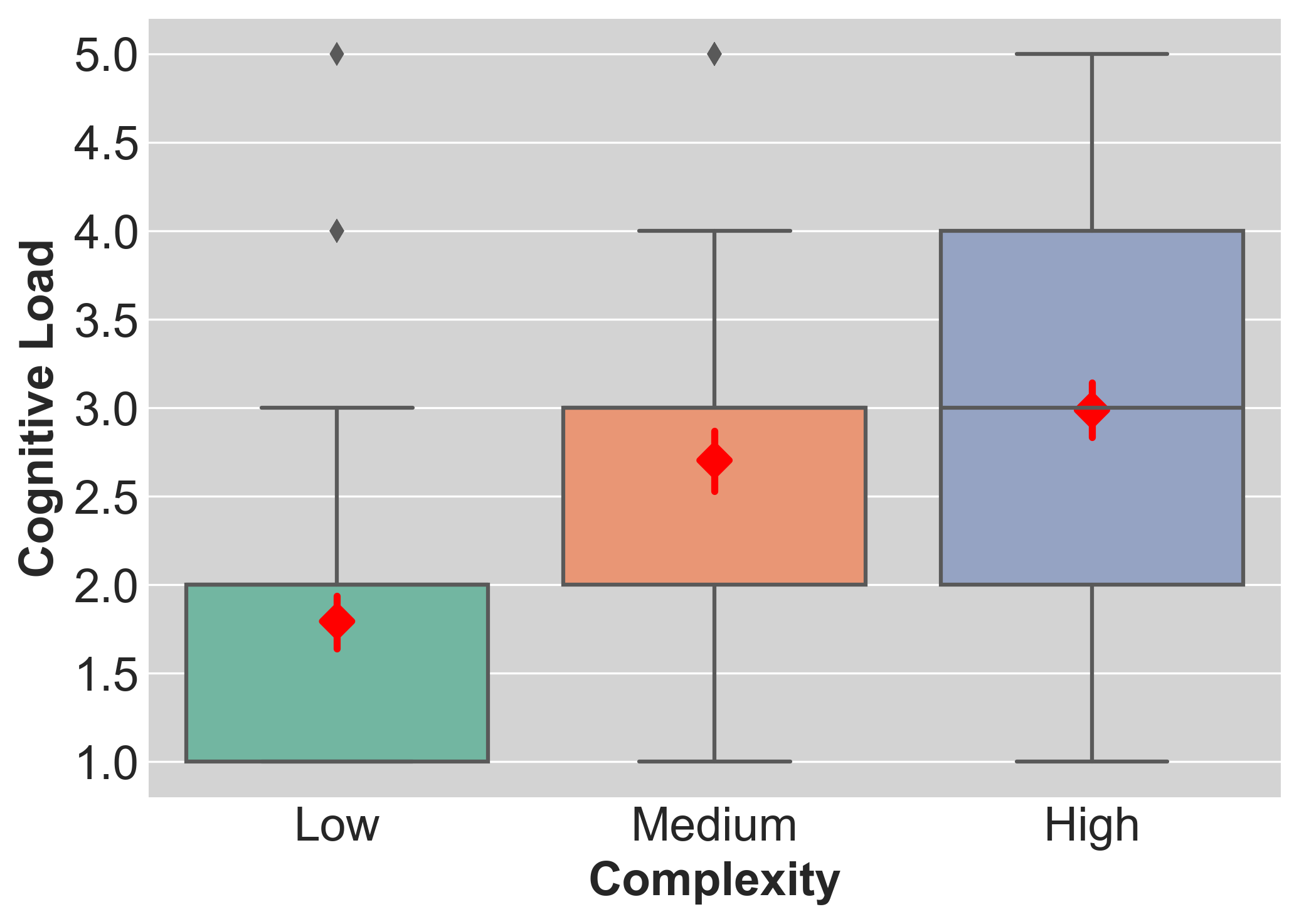}%
    }
    \hfill
    \subfloat[]{%
        \includegraphics[width=0.50\linewidth]{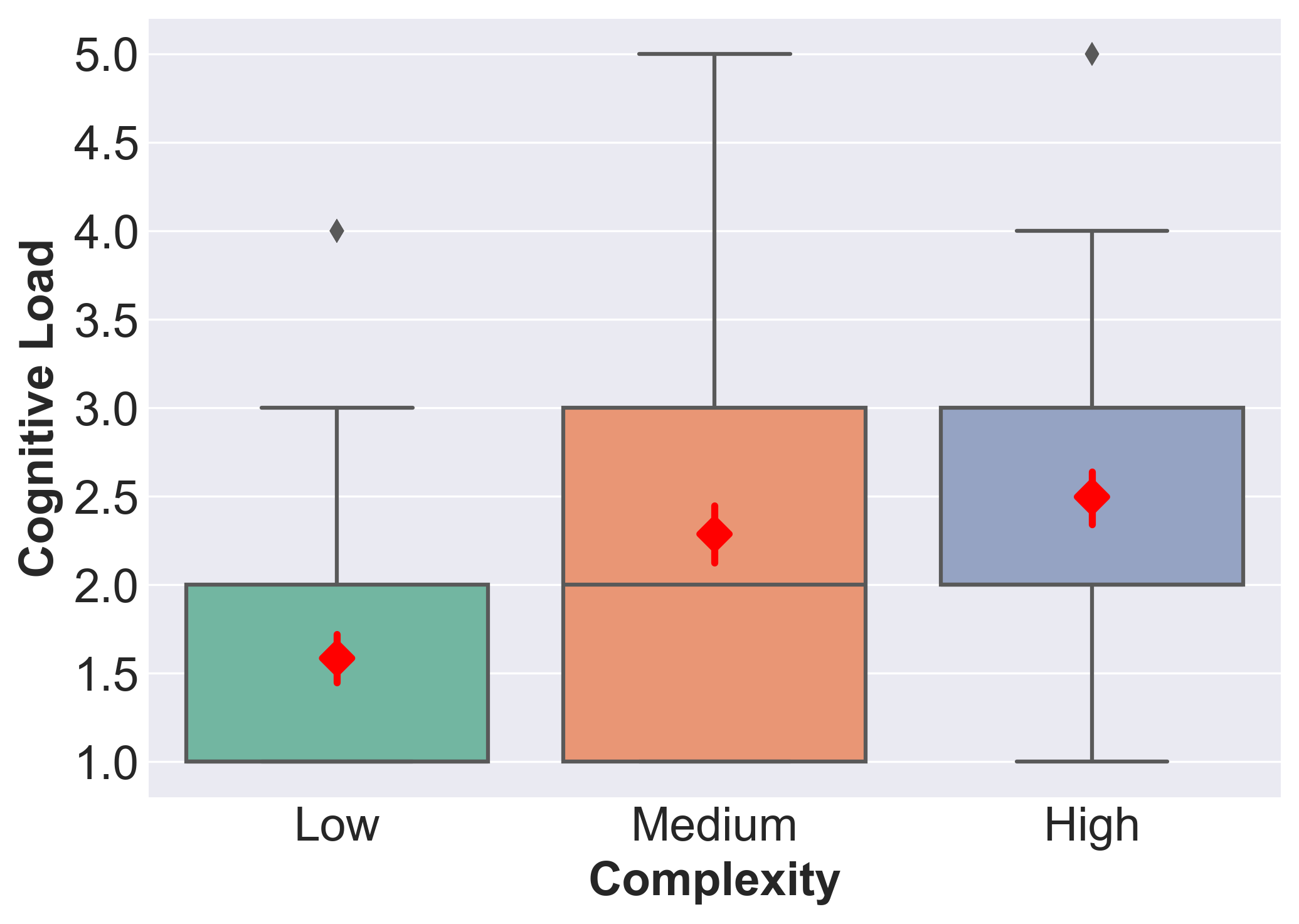}%
    }
    
    \subfloat[]{%
        \includegraphics[width=0.50\linewidth]{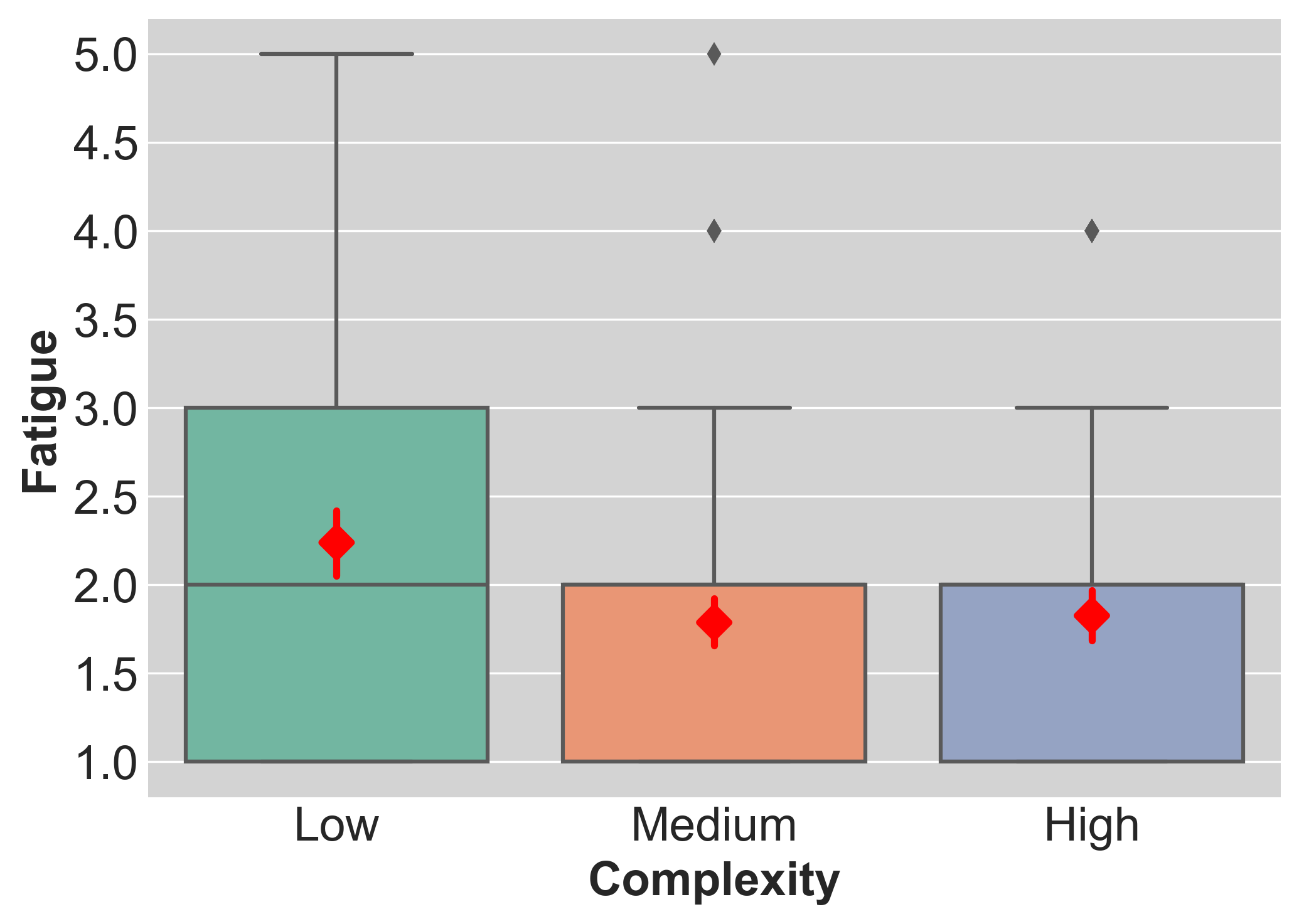}%
    }
    \hfill
    \subfloat[]{%
        \includegraphics[width=0.50\linewidth]{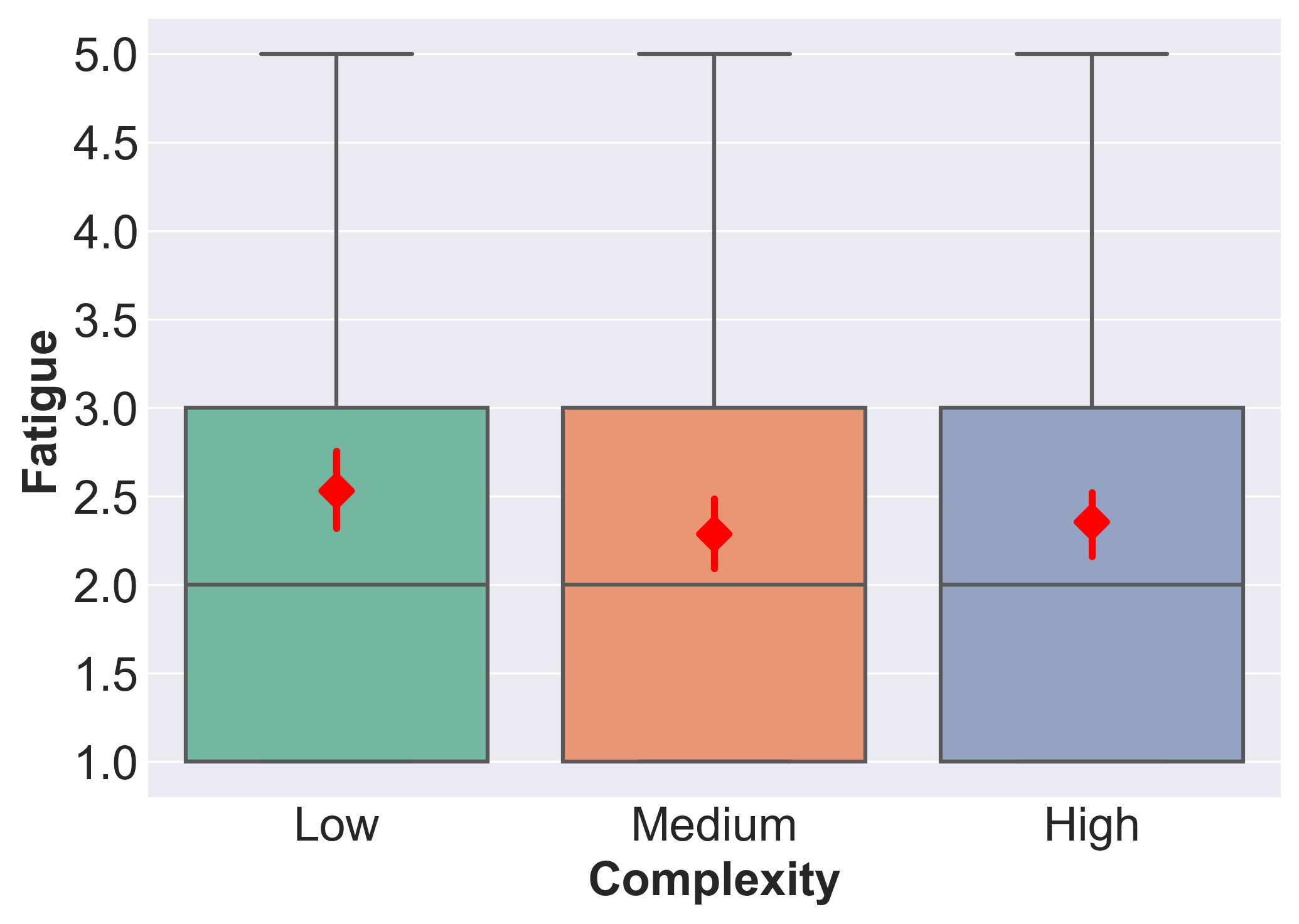}%
    }
    
    \subfloat[]{%
        \includegraphics[width=0.50\linewidth]{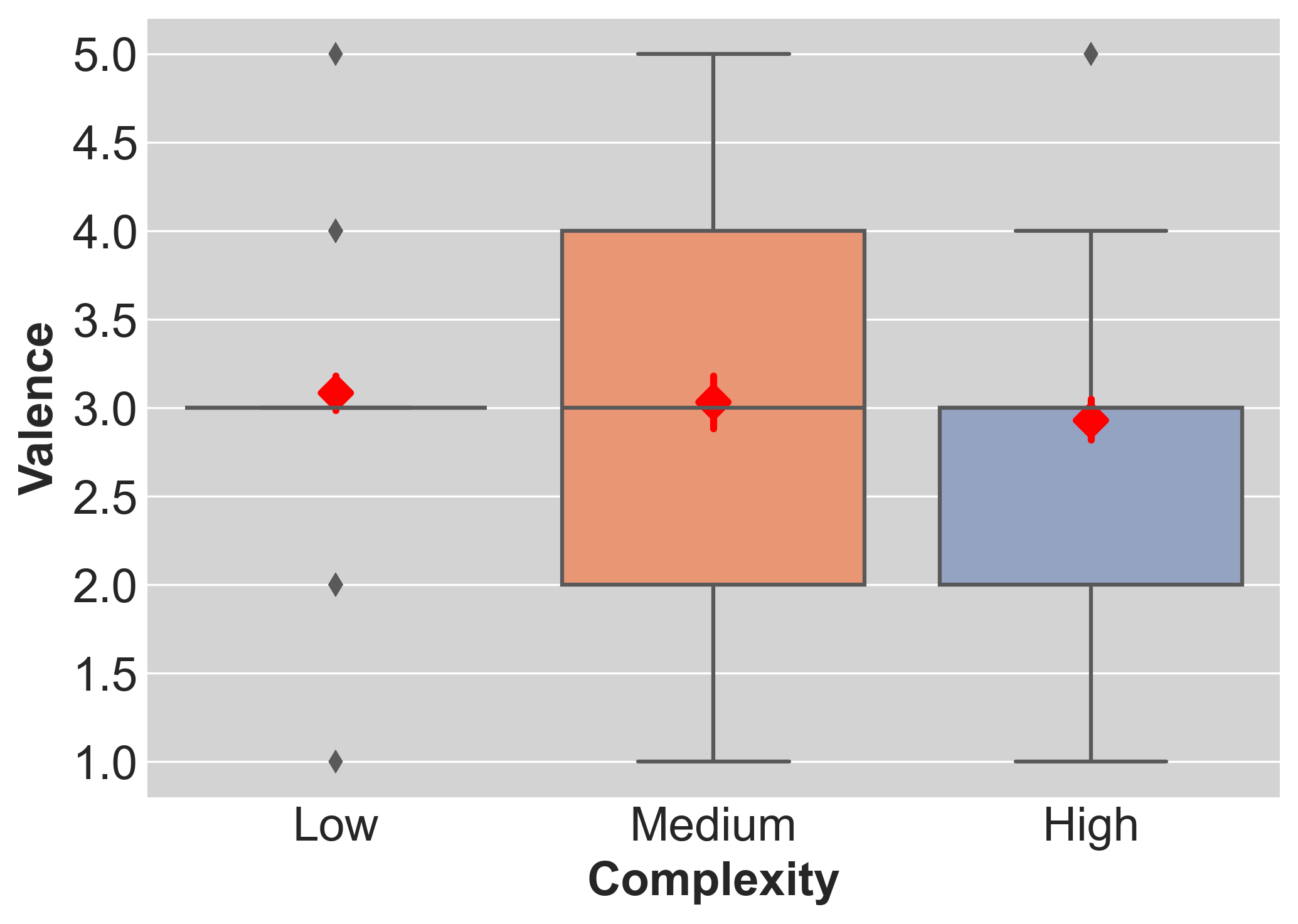}%
    }
    \hfill
    \subfloat[]{%
        \includegraphics[width=0.50\linewidth]{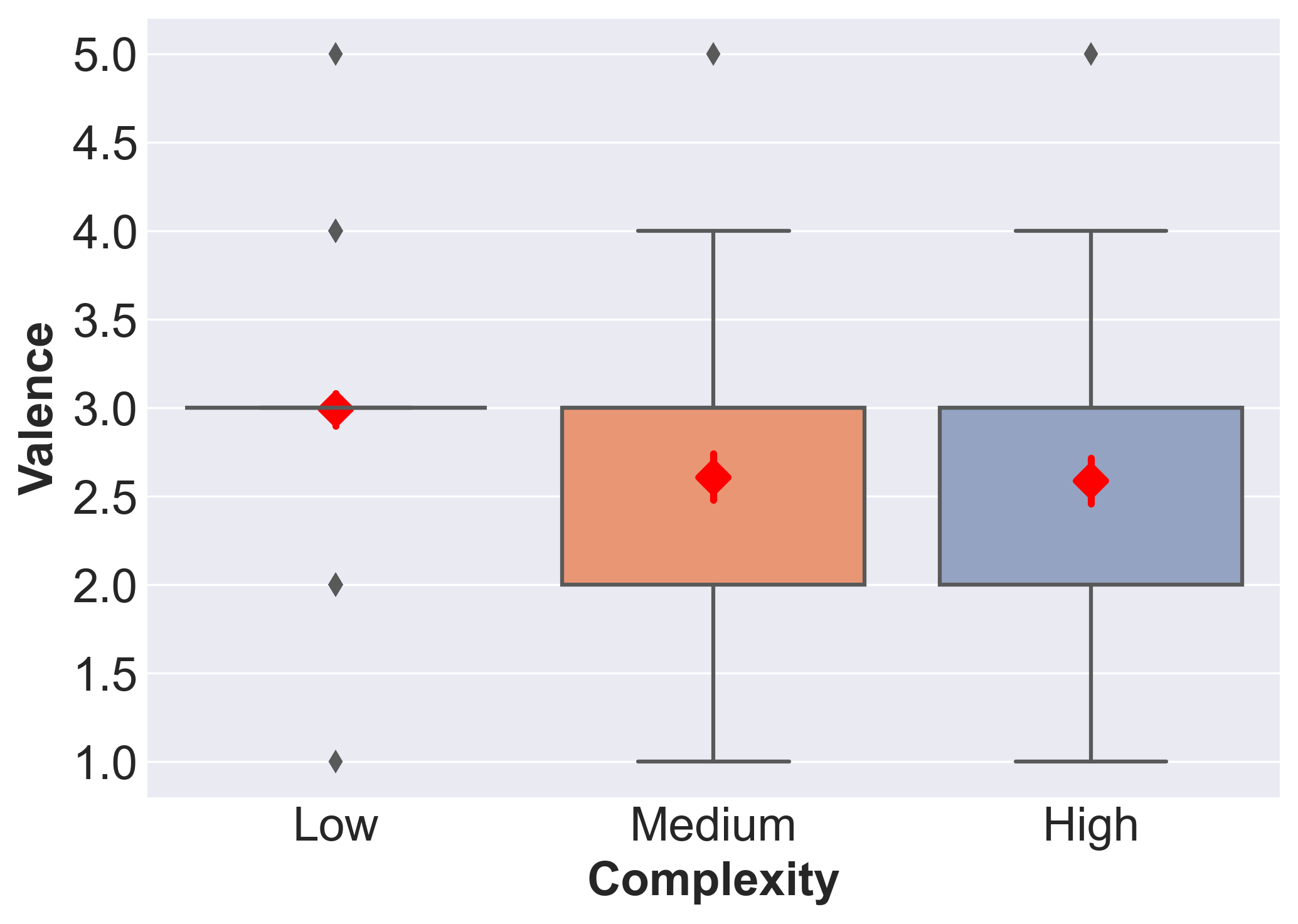}%
    }
    
    \subfloat[]{%
        \includegraphics[width=0.50\linewidth]{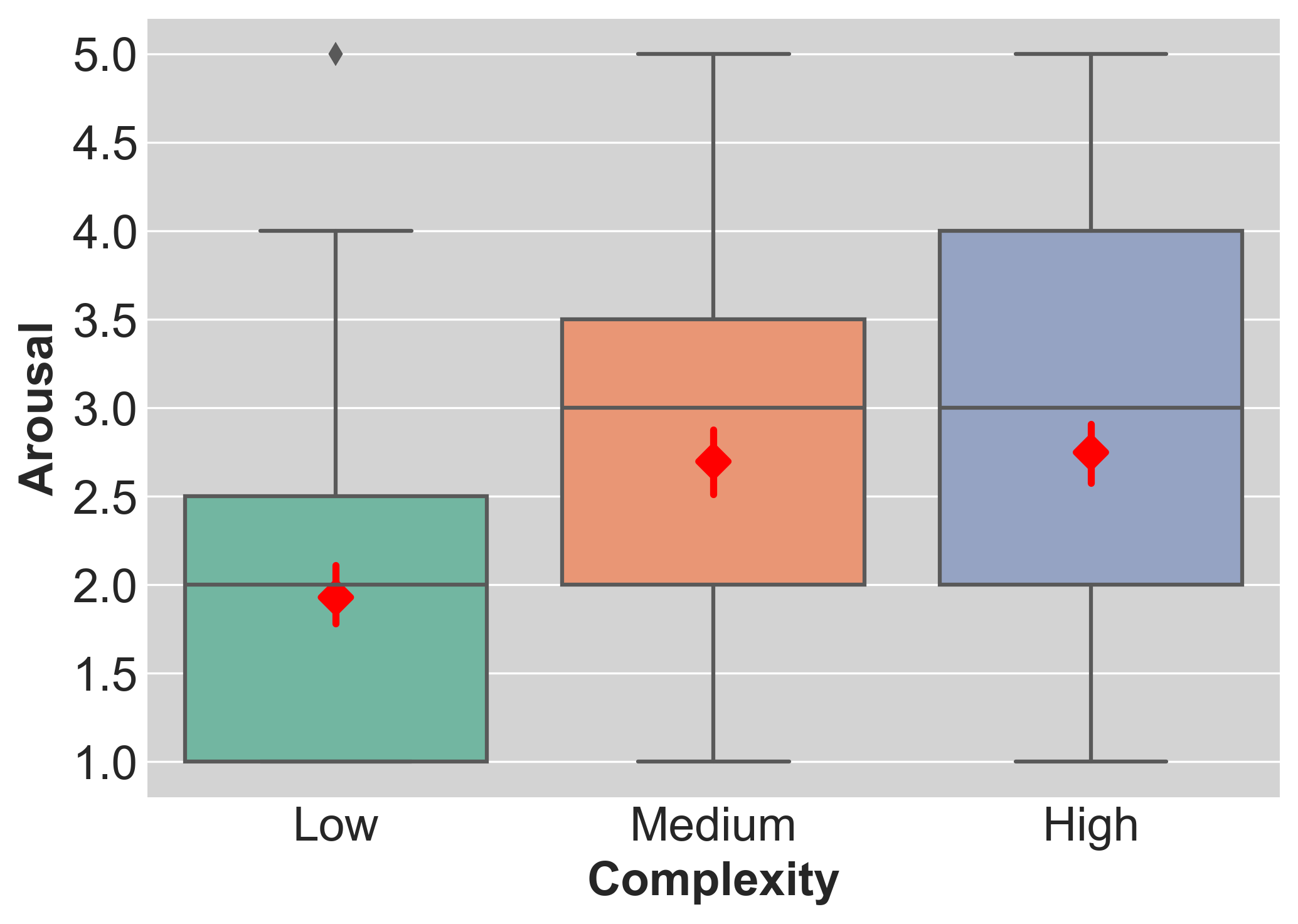}%
    }
    \hfill
    \subfloat[]{%
        \includegraphics[width=0.50\linewidth]{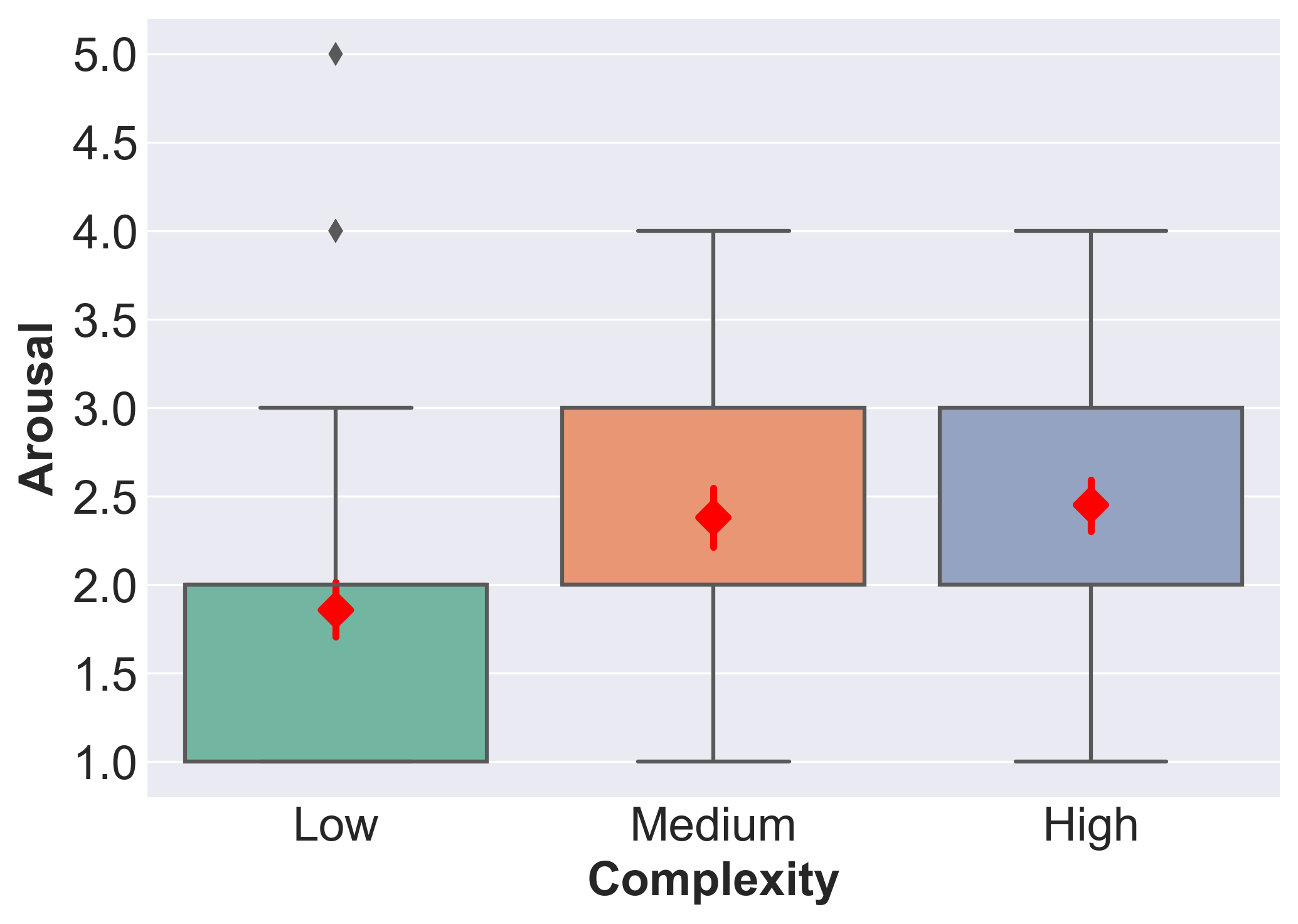}%
    }
    
    \caption{Complexity vs. mental states. The left plots are for active scenarios and the right plots are for autonomous scenarios.}
    \label{fig:complexity_vs_mental_states}
\end{figure}

Figure \ref{fig:complexity_vs_mental_states}, shows the relationship between task complexity levels and mental states in both driving conditions. Active scenarios are shown on the left in Figures \ref{fig:complexity_vs_mental_states} (a), (c), (e), and (g), and autonomous on the right in Figures \ref{fig:complexity_vs_mental_states} (b), (d), (f), and (h) across all participants. For cognitive load, (a) and (b) demonstrate that both active and autonomous driving scenarios exhibit the expected positive relationship with complexity. However, active driving demonstrates a steeper progression from low to high complexity, which is from 1.8 to 3.0, compared to autonomous driving, which is from 1.7 to 2.5. This indicates that while complexity affects cognitive demand in both conditions, the impact is more profound when participants actively control the vehicle versus passive monitoring. For fatigue, (c) and (d) show that low complexity levels produce a higher level of fatigue for both driving conditions. 
This is a phenomenon known as \textit{passive fatigue} \cite{korber2015vigilance,zhang2021electrophysiological}, which occurs in monotonous driving with little motor engagement and mainly passive monitoring. Monotonous and low-complexity autonomous scenarios therefore, tend to cause boredom-related passive fatigue, while higher complexity helps maintain alertness through increased visual and cognitive stimulation even without active control.
Similarly, during low-complexity sessions in the active driving scenarios, there is also minimal motor engagement and active control, which again induces passive fatigue. Both medium and high complexity levels show relatively lower fatigue for their respective driving scenarios.

For valence (e) and (f), we notice that high complexity sessions in both scenarios induce lower levels of valence compared to the medium and high complexity sessions. During high complexity, the participants need to be more vigilant while driving through situations such as traffic congestion to avoid pedestrians and accidents. These events likely cause lower valence during high complexity sessions. 
The opposite trend is observed for arousal (g) and (h) where high complexity sessions induce higher levels of arousal compared to low complexity sessions in both scenarios.  

\subsection{Valence vs Arousal}
\begin{figure}[t]
    \centering
    
    \subfloat[]{%
        \includegraphics[width=0.50\linewidth]{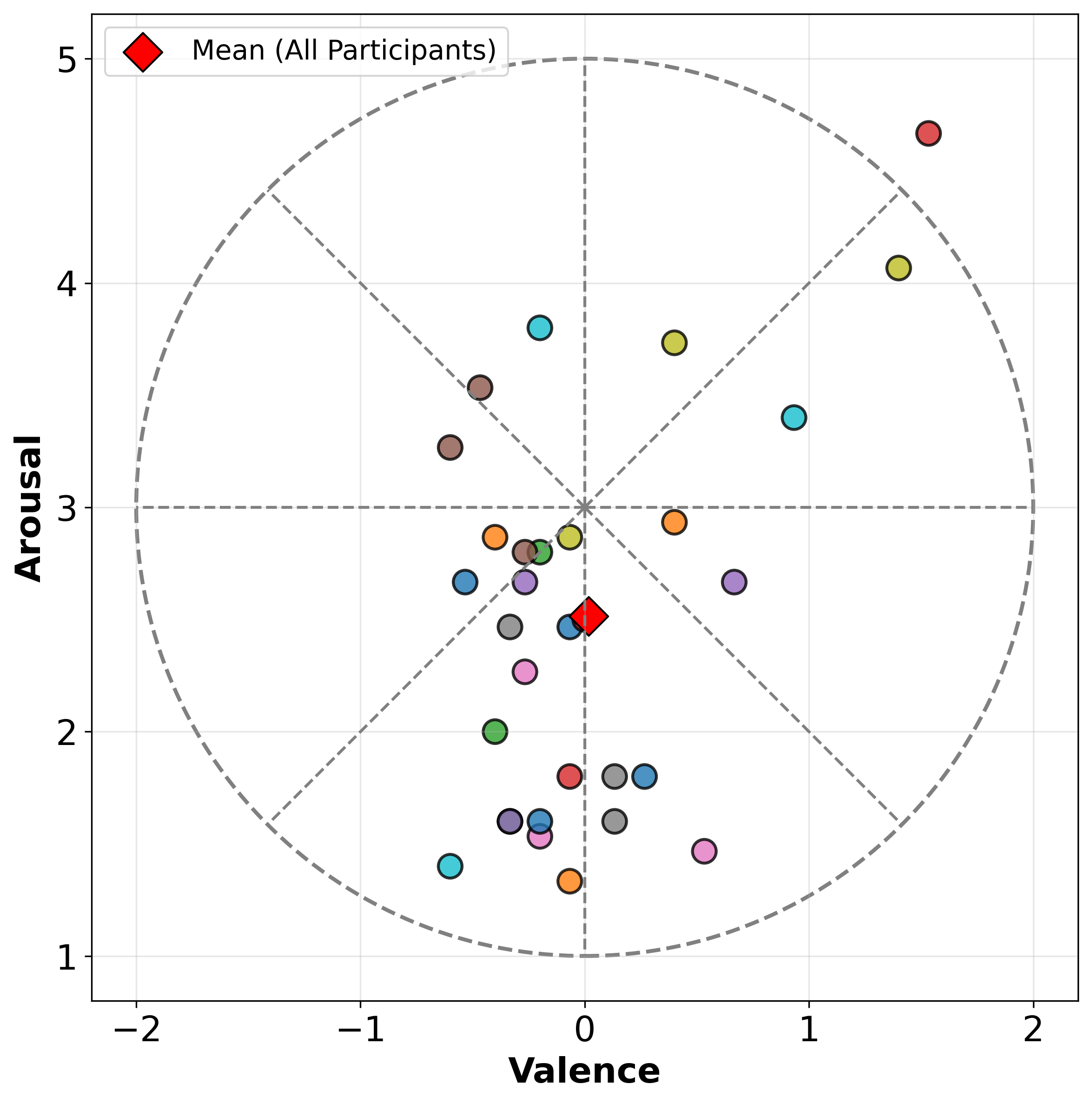}%
    }
    \hfill
    \subfloat[]{%
        \includegraphics[width=0.50\linewidth]{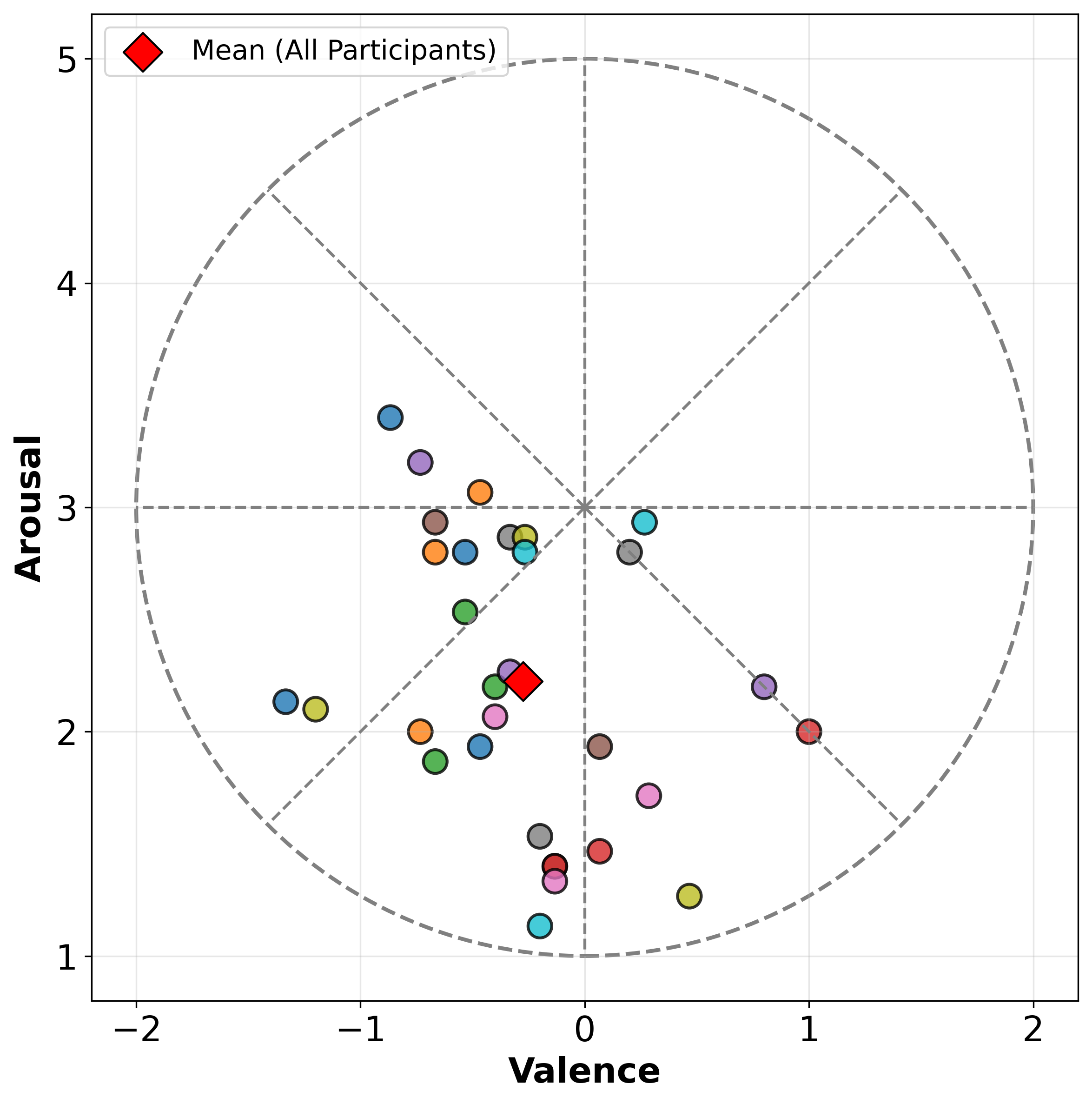}%
    }
    \caption{The average valence vs. arousal plots for all participants for active (left) and autonomous (right) driving scenarios.}
    \label{fig:valence_arousal}
\end{figure}

Figure \ref{fig:valence_arousal} shows the average valence and arousal of each participant on the circumplex model of affect \cite{russell1980circumplex}. We observe that for active scenarios (Figure \ref{fig:valence_arousal} (a)), on average, the participants experienced mostly low arousal and slightly positive valence, while 
for the autonomous scenario (Figure \ref{fig:valence_arousal} (b)), participants experienced lower arousal and valence. This could be attributed to higher excitement and a sense of accomplishment experienced during active driving. In exit interviews, for instance, several participants stated that they ``enjoyed'' the active driving while autonomous driving gave them ``no enjoyment'', made them ``bored'', or ``sleepy''. 





\subsection{Brain activity patterns}
\begin{figure}[t]
    \begin{center}
    \includegraphics[width=1\linewidth]{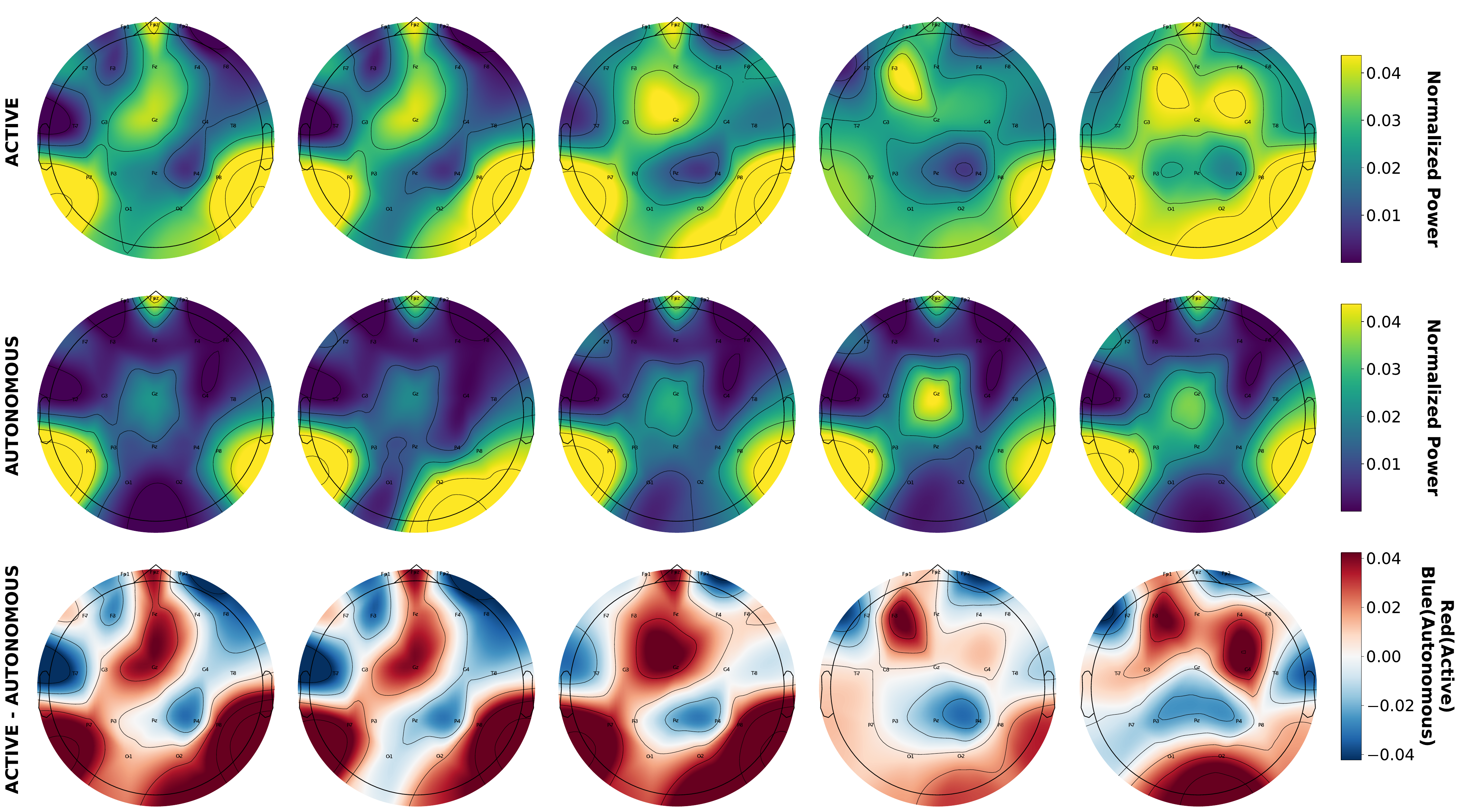}
    \caption{Brain activities in high complexity driving.}
    \label{fig:activity_high}
    \end{center}
\end{figure}

\begin{figure}[t]
    \begin{center}
    \includegraphics[width=1\linewidth]{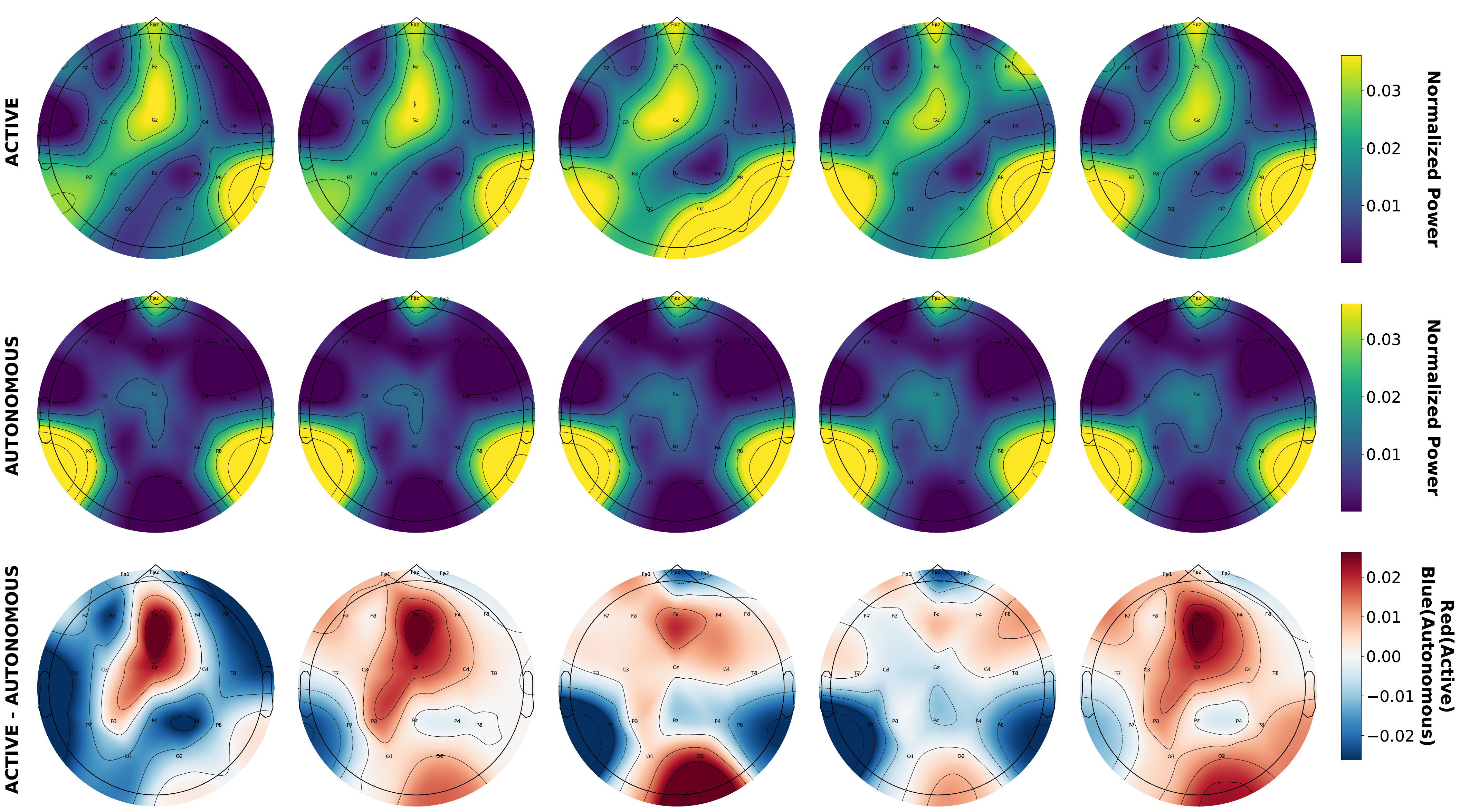}
    \caption{Brain activities in medium complexity driving.}
    \label{fig:activity_medium}
    \end{center}
\end{figure}

\begin{figure}[t]
    \begin{center}
    \includegraphics[width=1\linewidth]{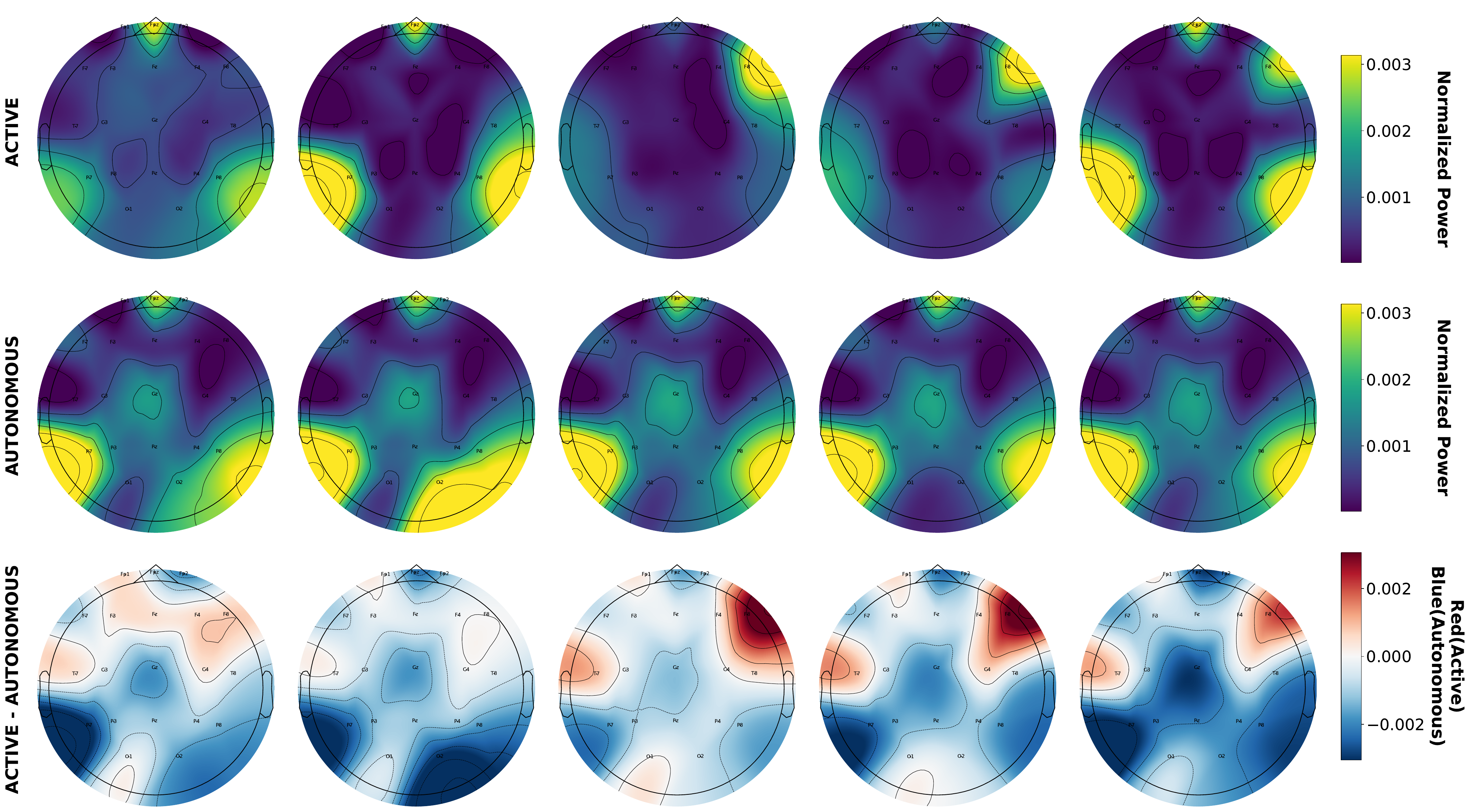}
    \caption{Brain activities in low complexity driving.}
    \label{fig:activity_low}
    \end{center}
\end{figure}

We construct topographic visualizations of EEG brain activity patterns across 5 temporal segments (0-2, 2-4, 4-6, 6-8, 8-10 minutes) for 3 different conditions: Active, Autonomous, and the difference between the two ($\Delta_{Active,Autonomous}$) in Figures \ref{fig:activity_high}, \ref{fig:activity_medium}, and \ref{fig:activity_low}. The top row in each figure represents the active scenario and the middle row represents the autonomous scenario, while the bottom row represents the difference between the two scenarios. The columns show the progression over time, each representing a 10-minute driving session. In Figure \ref{fig:activity_high}, the active scenario (top row) shows a relatively active pattern across all time segments, suggesting sustained engagement throughout the session. The brain activity is consistently high (yellow/light green regions) in frontal and temporal areas with moderate activity in the central regions. The frontal activation aligns with executive control and attention demands of active vehicle operation, and temporal lobe activity reflects auditory processing and spatial awareness. The Autonomous scenario (middle row) displays different patterns, with generally lower overall activity levels (blue/darker regions). The frontal regions show reduced activation, suggesting lower executive demands during passive monitoring. In the $\Delta_{Active,Autonomous}$ difference maps (bottom row), red indicates higher activity in active driving and blue indicates higher activity in autonomous driving. The early segments (0-4 minutes) show strong red regions in frontal and parietal areas, indicating notably higher activity during active compared to autonomous driving. The blue regions become more prominent during 6-10 minutes, suggesting that autonomous driving may eventually lead to increased activity in certain brain regions. This increase could be due to decreased vigilance, mind drifting, or passive fatigue as participants adapt to the passive observation role.

The Active condition (top row) in Figure \ref{fig:activity_medium} shows moderate and consistent brain activity patterns with less intense frontal and temporal activation compared to high-complexity driving. This suggests that medium-complexity driving requires sustained but less intensive cognitive engagement than high-complexity scenarios. The Autonomous condition (middle row) maintains the overall low activity pattern similar to that of high-complexity. However, the temporal areas show consistent yellow activation across all time segments, which is more stable in medium complexity compared to the variable patterns observed in high complexity autonomous driving. The central regions remain consistently low in activation, suggesting less motor engagement. The early segments of $\Delta_{Active,Autonomous}$ maps (bottom row) show dark red in the frontal regions, indicating higher active driving engagement, while the middle segments are more balanced, and the last segment is dark red again.

The topography maps for low-complexity scenarios shown in Figure \ref{fig:activity_low} demonstrate how reduced task demand fundamentally changes the brain engagement patterns. The Active condition (top row) shows some temporal activity and moderate activity in frontal regions. This pattern suggests that low-complexity active driving may engage some attention and motor control, but as low-complexity driving is monotonous and requires comparably less motor engagement, the brain activity shows reduced activation compared to high and medium complexity levels. 
The Autonomous condition (middle row) shows some occipital activation and consistent temporal activation, suggesting auditory and visual processing of the driving environment.


\subsection{Classifier Performance}
As discussed in Section~\ref{section:data_processing}, we trained a Transformer-based classifier to classify driver mental states using a LOSO cross-validation strategy. Our objective is to evaluate whether EEG-based mental state classification remains reliable across both active and autonomous driving conditions and to examine how pre-training and fine-tuning influence model performance in these two scenarios.

The results for the active and autonomous driving scenarios are shown in Tables \ref{table:active_classification} and \ref{table:autonomous_classification}. For the active sessions trained without pre-training, among the 4 classification tasks, valence shows the best results with an F1 of 60.45\% and an accuracy of 77.03\%. This is followed by fatigue with an F1 and accuracy of 60.35\% and 78.35\%, respectively. For autonomous sessions, trained without pre-training, the best results are achieved for arousal, with an F1 and accuracy of 62.71\% and 69.60\%. Moreover, cognitive load shows the second-best performance with an F1 and accuracy of 69.60\% and 60.71\%.

For active driving scenarios, the frozen pretrained features show modest improvements over training from scratch, with F1 and accuracy gains of 1.82\% and 0.26\% for fatigue, 0.87\% and 0.23\% for valence, and 0.29\% and 1.29\% for arousal classification. However, we see a decline in performance for cognitive load with a change in F1 and accuracy of 3.99\% and 4.64\%. The frozen approach shows moderate improvements in autonomous driving for fatigue and arousal classification. We observe a 2.09\% improvement in F1 and a 2.36\% improvement in accuracy for fatigue, while for arousal, the performance increases by 1.76\% and 0.98\%, respectively. Finally, decreases in cognitive load and valence are observed with changes in F1 of 0.47\% and 1.5\%, respectively.

The results demonstrate that fine-tuning the entire network provides substantial benefits for EEG-based mental state classification across both active and autonomous driving conditions. For active driving scenarios, the fine-tuning shows an improvement of F1 of 2.02\%, 5.10\%, 7.95\% and 2.7\% for cognitive load, fatigue, valence, and arousal, respectively. For the autonomous scenario, there are improvements of 1.75\%, 2.65\%, 0.14\% and 3.34\% for the mental states, respectively.

Overall, the results show that EEG-based mental state classification remains feasible and stable across both active and autonomous driving scenarios. While the mental states exhibit small performance differences between the two conditions, the absence of motor engagement in autonomous driving does not lead to a substantial decrease in classification performance. These findings highlight the potential for developing dedicated EEG-based monitoring systems made specifically for autonomous driving contexts.

\begin{table*}[htbp]
\centering
\caption{Classification results in \textit{active} driving.}
\begin{tabular}{l|c|c|c|c|c|c}
\hline
\multirow{2}{*}{\textbf{Mental States}} & \multicolumn{2}{c|}{\textbf{Without pre-training}} & \multicolumn{2}{c}{\textbf{Frozen}} & \multicolumn{2}{c}{\textbf{Unfrozen}} \\
\cline{2-7}
 & \textbf{Accuracy} & \textbf{F1} & \textbf{Accuracy} & \textbf{F1} & \textbf{Accuracy} & \textbf{F1} \\
\hline
Cognitive Load & 70.68 (12.70) & 60.62 (13.94) & 66.04 (18.26) & 56.63 (17.13) & 70.99 (11.18) & 62.64 (13.15) \\
Fatigue & 78.35 (14.97) & 60.35 (25.80) & 78.61 (22.63) & 62.17 (23.81) & 77.60 (20.76) & 65.45 (20.15) \\
Valence & 77.03 (20.41) & 60.45 (25.48) & 77.26 (20.43) & 61.32 (23.15) & 77.29 (18.44) & 68.40 (19.14) \\
Arousal & 70.42 (16.41) & 56.24 (13.97) & 71.71 (20.65) & 56.53 (18.46) & 74.65 (19.87) & 58.94 (17.35) \\
\hline
\end{tabular}
\label{table:active_classification}
\end{table*}

\begin{table*}[htbp]
\centering
\caption{Classification results in \textit{autonomous} driving.}
\begin{tabular}{l|c|c|c|c|c|c}
\hline
\multirow{2}{*}{\textbf{Mental States}} & \multicolumn{2}{c|}{\textbf{Without pre-training}} & \multicolumn{2}{c}{\textbf{Frozen}} & \multicolumn{2}{c}{\textbf{Unfrozen}} \\
\cline{2-7}
 & \textbf{Accuracy} & \textbf{F1} & \textbf{Accuracy} & \textbf{F1} & \textbf{Accuracy} & \textbf{F1} \\
\hline
Cognitive Load & 69.60 (21.48) & 60.71 (22.49) & 70.59 (22.80) & 60.24 (19.51) & 71.91 (19.08) & 62.46 (12.26) \\
Fatigue & 69.88 (19.08) & 58.35 (19.56) & 72.24 (18.43) & 60.44 (17.62) & 76.78 (20.02) & 61.00 (18.59) \\
Valence & 68.20 (16.32) & 59.92 (18.63) & 69.26 (15.43) & 58.42 (14.64) & 70.35 (13.29) & 60.06 (13.81) \\
Arousal & 69.60 (19.65) & 62.71 (19.05) & 70.58 (20.73) & 64.47 (19.19) & 77.78 (18.45) & 66.05 (17.97) \\
\hline
\end{tabular}
\label{table:autonomous_classification}
\end{table*}

\subsection{Domain shift}
\begin{figure}
    \begin{center}
    \includegraphics[width=0.65\linewidth]{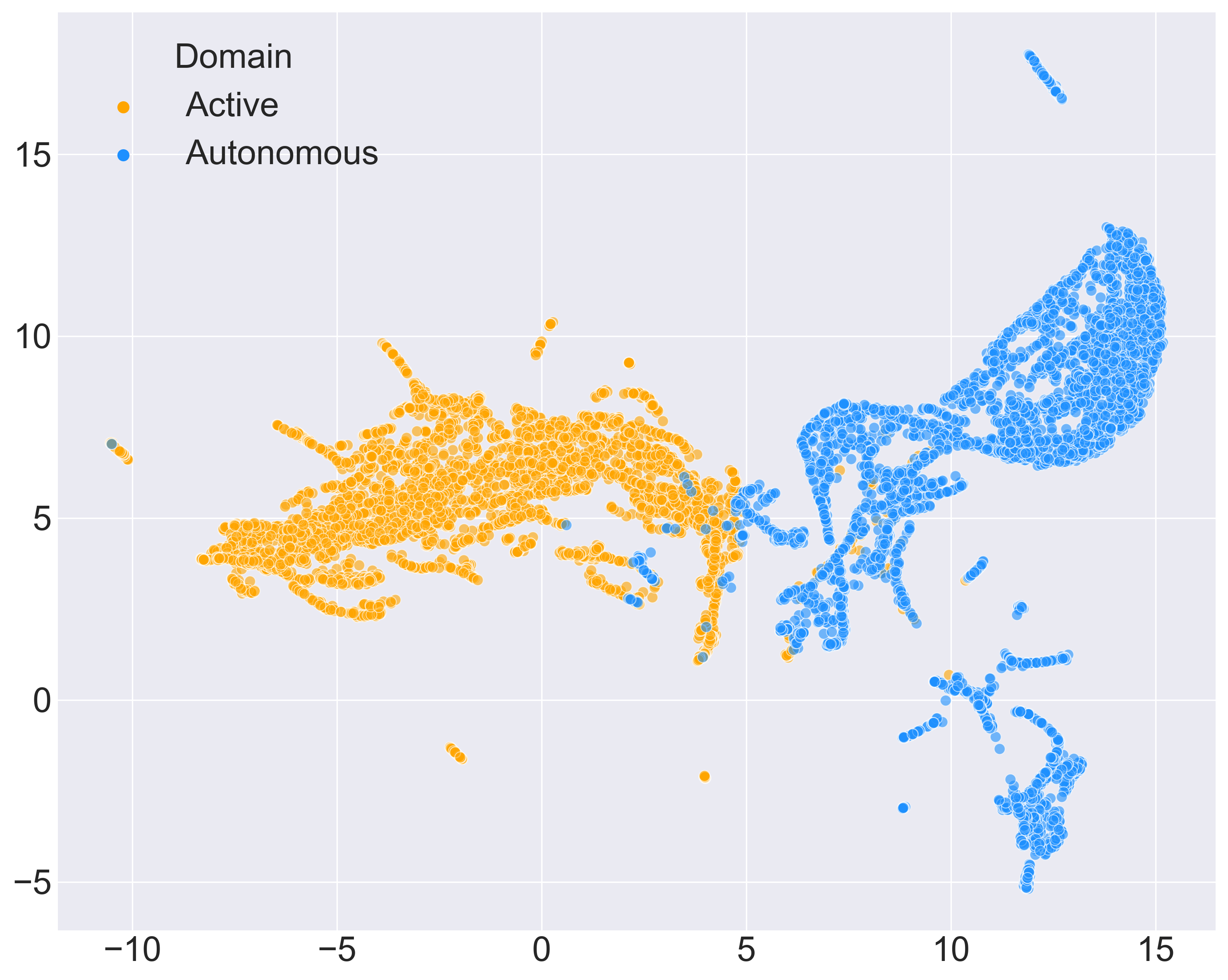}
    \caption{UMAP visualization showing the distribution shift between active and autonomous driving scenarios across all participants. 
    }
    \label{fig:distribution_shift}
    \end{center}
\end{figure}

To examine how active and autonomous driving conditions are distributed within the representation space learned during pre-training, we apply Uniform Manifold Approximation and Projection (UMAP) \cite{mcinnes2018umap} to the learned embedding of both driving scenarios. This results in a low-dimensional projection that preserves both local and global structures, enabling us to visualize and compare the distribution of representations and identify potential domain shifts between the two driving scenarios. The results are presented in Figure \ref{fig:distribution_shift}, 
where a clear domain shift between active and autonomous driving conditions is observed. To obtain the embeddings, we use an encoder trained on SEED and SEED-IV datasets using masked frequency band prediction.
The distinct clustering patterns in the learned embedding space are shown in blue for active and orange for autonomous data. This spatial separation demonstrates that despite using identical EEG pre-processing pipelines and electrode configurations, the underlying neural activity patterns during active versus autonomous driving are largely distinct. 
This domain shift reflects the fundamental differences in mental states 
between the two driving conditions. During active driving, participants must continuously monitor the environment, make rapid decisions, and execute motor responses, resulting in specific patterns of brain activity. This causes increased cognitive load and sensorimotor engagement. On the contrary, autonomous driving allows for more passive observation and reduced active control, causing potentially lower vigilance states and different attention patterns.

We also conduct an experiment with cross-session training to show how the domain shift impacts the results. Table \ref{table:domain_shift} presents the results for training the model on active scenario and testing on autonomous scenario and vice versa. Comparing the result with Table \ref{table:active_classification} and \ref{table:autonomous_classification}, we can see that there is a decrease in performance in the cross-session training. Specifically, in active scenario the F1 decreases by 6.30\%, 4.91\%, 6.62\%, and 4.09\% for cognitive load, fatigue, valence, and arousal, respectively. In the autonomous scenario, there is a decrease of 1.76\% in cognitive load, 9.34\% in fatigue, 7.10\% valence, and 3.40\% arousal. This decrease further confirms the domain shift between active and autonomous driving conditions.

    
    

\begin{figure}
    \begin{center}
    \includegraphics[width=1\linewidth]{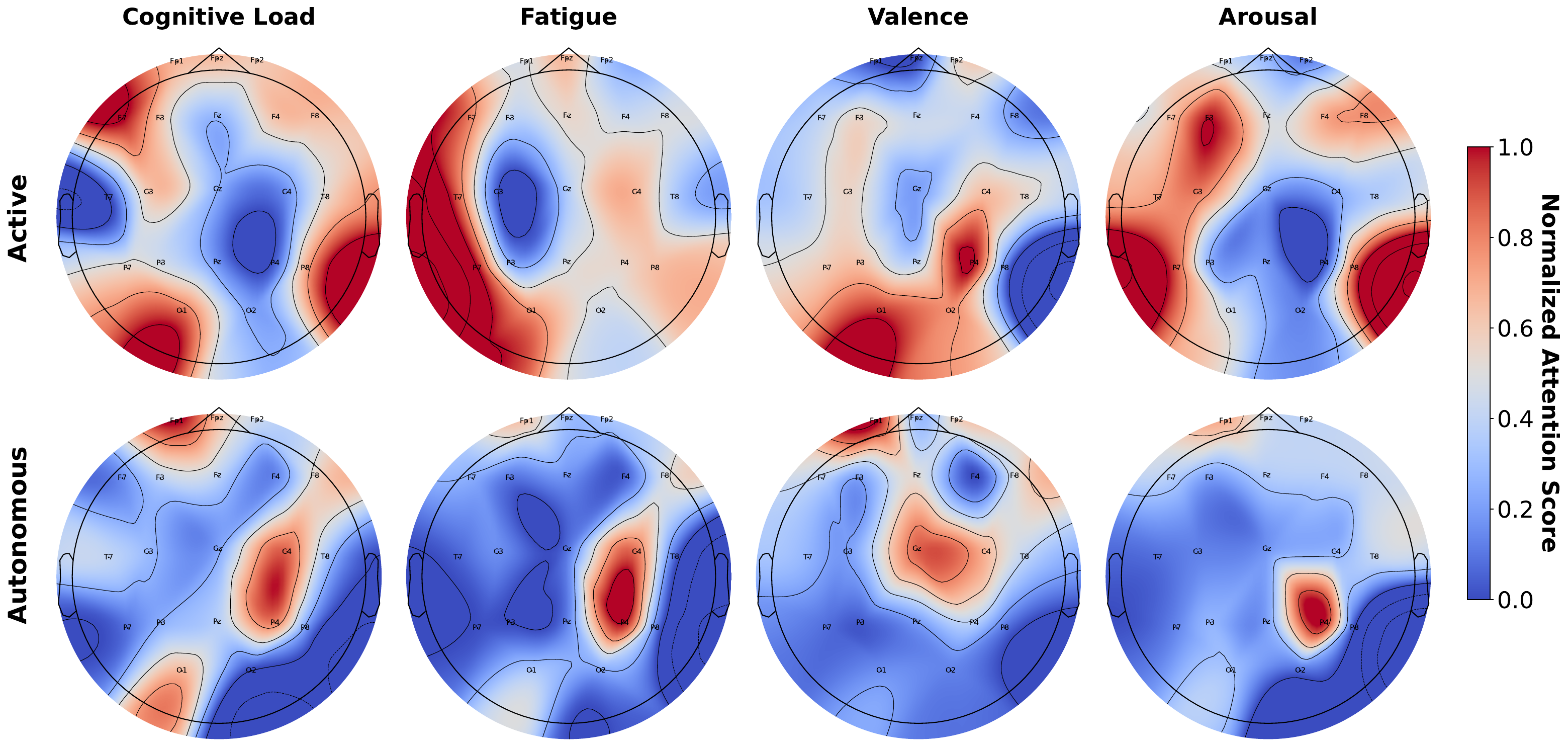}
    \caption{Channel attention obtained from the Transformer model.}
    \label{fig:channel_attention}
    \end{center}
\end{figure}

\begin{table*}[htbp]
\centering
\caption{Cross-scenario training results.}
\begin{tabular}{l|c|c|c|c}
\hline
\multirow{2}{*}{\textbf{Mental States}} & \multicolumn{2}{c}{\textbf{Train on Autonomous Test on Active}} & \multicolumn{2}{c}{\textbf{Train on Active Test on Autonomous}}  \\
\cline{2-5}
 & \textbf{Accuracy} & \textbf{F1 score} & \textbf{Accuracy} & \textbf{F1 score} \\
\hline
Cognitive Load & 63.98 (14.98) &  54.32 (13.24) &  65.07 (19.31) &  58.95 (17.61) \\
Fatigue &  74.52 (16.79) &  55.44 (23.36) & 65.97 (21.67) &  49.01 (20.72)  \\
Valence &  72.85 (17.51) &  53.83 (24.48) &  68.76 (14.42) &  52.82 (20.71) \\
Arousal &  69.95 (17.67) &  52.15 (16.75) &  70.75 (23.64) &  59.31 (20.73)  \\
\hline
\end{tabular}
\label{table:domain_shift}
\end{table*}

\subsection{Channel-wise attention}

To better understand how different neural regions or channels contribute to mental-state processing during driving, we group EEG electrodes into major cortical areas that correspond to regions of the brain. We categorize electrodes as follows: prefrontal cortex (Fp1, Fp2, Fpz), frontal cortex (F3, F4, Fz, F7, F8), central cortex (C3, C4, Cz), temporal cortex (T7, T8), parietal cortex (P3, P4, Pz, P7, P8), and occipital cortex (O1, O2). Prior research have shown that these regions correspond to different cognitive, affective, motor or sensory functions. The prefrontal cortex is associated with executive control, decision-making, and regulation of attention \cite{gazzaley2007unifying,krawczyk2002contributions,knight1994attention}. The frontal cortex plays a key role in working memory, cognitive load, and voluntary motor planning \cite{buchsbaum2004frontal}. The central cortex, located over the sensorimotor strip, is involved in motor preparation and movement execution \cite{sanes2000plasticity}. The temporal cortex supports auditory processing and affective evaluation \cite{schirmer2017temporal}. The parietal cortex is linked to visuospatial processing, attention allocation, and integration of sensory information \cite{gottlieb2010spatial, culham2006human}. 
Finally, the occipital cortex is responsible for visual perception and processing \cite{blihar2020systematic}.

To identify which EEG channels contribute most significantly to mental state classification in our data, we employ a channel-wise attention analysis using the inherent attention mechanism of the trained Transformer model. 
The normalized channel importance plot for each mental state and each scenario, averaged across all participants, is shown in Figure \ref{fig:channel_attention}, where blue to red indicates least to most important.
For cognitive load classification, in active driving, our analysis reveals F7 (left frontal) as the most important channel, followed by O1 (left occipital) and P8 (right parietal). During the autonomous scenario, Fp1 (left prefrontal cortex) is found to be the most important, closely followed by C4 (right central). 
For fatigue classification, in active driving, P7 (left parietal) region is the leading channel, followed by T7 (left temporal) and F7 (left frontal). In the autonomous scenario, P4 (right parietal) is most important followed by C4 (right central) and Fp1 (left prefrontal). 
For valence classification, in the active scenario, the most attention is put into P4 (right parietal) followed by left and right occipital electrodes O1 and O2. 
We observe that Fp1 (left prefrontal cortex) is the most important channel in the autonomous scenario for detecting valence. 
For arousal classification, in the active driving scenario, F3 (left frontal), P7 and P8 (left and right parietal) regions are most important. 
In the autonomous scenario, P4 (right parietal) is the most important channel, followed by Fp1 (left prefrontal cortex) and Fpz (prefrontal cortex). 
This analysis shows that the Transformer’s channel-wise attention is spatially structured and aligns with established functional roles of major cortical regions. Frontal and prefrontal channels associated with executive control, attentional regulation, and motor planning are most informative for cognitive load, valence, and arousal. Parietal and occipital regions related to visuospatial and visual processing contribute strongly to valence, arousal, and fatigue, particularly under visually demanding conditions. Central and temporal regions exhibit task-dependent contributions linked to motor engagement and affective processing. Overall, the learned importance patterns reflect systematic and state-specific shifts in regional EEG relevance across active and autonomous driving.

\section{Conclusion and Future Work}
In this work, we provide the first EEG-based comparison of mental states during active and autonomous driving and show that, although both scenarios share identical visual stimuli and task complexities, participants experience different brain activities when comparing the two driving modes. We carried out temporal, complexity-wise, and statistical analyses to reveal the differences between the two driving conditions. Our transfer-learning experiment and UMAP visualization confirm a clear distribution shift between active and autonomous EEG signals. Despite the absence of motor-related brain activity in autonomous driving, our neural network successfully detects cognitive load, fatigue, valence, and arousal, demonstrating that mental state monitoring remains reliable even in passive driving environments. Together, these findings highlight that active and autonomous driving rely on different functional contributions from major brain regions, even under identical visual and task demands. 
Moreover, the distinct brain activation and channel-wise attention patterns observed across mental states and driving scenarios indicate engagement of different neural regions as drivers’ mental states vary.
These differences help explain the observed distribution shift and emphasize that mental states in autonomous vehicles cannot be inferred simply by extrapolating from active driving data. Ultimately, these findings highlight the importance of collecting specific data for autonomous vehicles to design safe and robust driver-state monitoring systems.


    

    


\bibliographystyle{IEEEtran}
\bibliography{refs}

\end{document}